\documentclass[reprint,twocolumn,secnumarabic,amssymb, nobibnotes, aps, pre,showkeys]{revtex4-2}
\usepackage{bm,chemformula,graphicx,titlesec}
\usepackage[unicode]{hyperref}
\hypersetup{colorlinks= true,
	citecolor= blue,
	linkcolor= red,
	urlcolor=magenta,
	filecolor=cyan,
	linkbordercolor={1 0 0},
	citebordercolor={0 1 0},
	urlbordercolor={0 1 1},
} 
\usepackage{cleveref}
\usepackage[mathlines]{lineno}
\usepackage[section]{placeins}
\usepackage{tikz}
\usepackage{float,wrapfig}
\usepackage{microtype}
\usepackage[tight]{subfigure}
\DeclareUnicodeCharacter{2212}{-}
\begin{document}
	\preprint{APS/123-QED}
	\title{Nonequilibrium thermodynamic signatures of collective dynamical states around chimera in a chemical reaction network}
	\author{Premashis Kumar}
	\email{pkmanager@bose.res.in}
	\affiliation{S. N. Bose National Centre For Basic Sciences, Block-JD, Sector-III, Salt Lake, Kolkata 700 106, India}
	\author{Gautam Gangopadhyay}
	\email{gautam@bose.res.in}
	\affiliation{S. N. Bose National Centre For Basic Sciences, Block-JD, Sector-III, Salt Lake, Kolkata 700 106, India}
	\date{\today}
	\begin{abstract}
		Different dynamical states ranging from coherent, incoherent to chimera, multichimera, and related transitions are addressed in a globally coupled nonlinear continuum chemical oscillator system by implementing a modified complex Ginzburg-Landau equation. Besides dynamical identifications of observed states using standard qualitative metrics, we systematically acquire nonequilibrium thermodynamic characterizations of these states obtained via coupling parameters. The nonconservative work profiles in collective dynamics qualitatively reflect the time-integrated concentration of the activator, and the majority of the nonconservative work contributes to the entropy production over the spatial dimension. It is illustrated that the evolution of spatial entropy production and semigrand Gibbs free energy profiles associated with each state are connected yet completely out of phase, and these thermodynamic signatures are extensively elaborated to shed light on the exclusiveness and similarities of these states. {Moreover, a relationship between the proper nonequilibrium thermodynamic potential and the variance of activator concentration is established by exhibiting both quantitative and qualitative similarities between a Fano factor-like entity, derived from the activator concentration, and the Kullback-Leibler divergence associated with the transition from a nonequilibrium homogeneous state to an inhomogeneous state.}  Quantifying the thermodynamic costs for collective dynamical states would aid in efficiently controlling, manipulating, and sustaining such states to explore the real-world relevance and applications of these states.           
	\end{abstract}                  
	\maketitle
	\section{INTRODUCTION}\label{sec1}
	Chimera~\cite{kuramoto2002coexistence, strogatz}, a subtle and counterintuitive spatiotemporal pattern with spatial coexistence of coherent and incoherent states, is one of the most debated topics in collective dynamics of systems and is reported in diverse theoretical frameworks~\cite{sethiasen0, spiralchimera, motter2010spontaneous, turbulent, Lainglocal, quantumchimera, Selfpropelled} and experimental settings ranging from optical configuration~\cite{hagerstrom2012experimental}, nonlocally coupled photosensitive Belousov-Zhabotinsky (BZ) chemical oscillators~\cite{Tinsleynature2012} to network of mechanical oscillators~\cite{Martens10563} and laser array~\cite{PhysRevE.91.040901, larger2015laser}. Although initially chimeras were only attributed to the phase dynamics of identical oscillator collection, findings of amplitude-mediated~\cite{sethiasen1, schmidtglobal, sethiasen} and amplitude chimeras~\cite{amplitudechimera} expand their existence in more general situations. For the chemical reaction systems, diverse chimeras have been illustrated in nonlocally coupled chemical oscillators~\cite{Tinsleynature2012, chemicaltheoretical, wickramasinghe2013spatially, nkomo2016, totz2018spiral}. Lately, chimera states based on the amplitude dynamics of a simple prototypical continuum chemical oscillator system have also been generated by implementing a global coupling scheme~\cite{pkgg3}. Despite being a counterintuitive state, chimeras are sought to be a natural link between the coherence and incoherence state~\cite{naturallink}. In the coupled nonlinear system of identical oscillators, the transition from a coherent state to a completely incoherent state or from cluster states to coherent behavior can be mediated via the chimera state~\cite{coherenceincoherencetransition, prerequisite, strongchimera}. 
	
	Until now, the main focus of the investigation and the arguments related to chimera and associated transitions or identifying different states along transitions are purely dynamics-oriented.{ In the context of global coupling, different kinds of complex bifurcation scenarios associated with various states in a coupled system, mechanisms of creation and destruction of torus motion, chaos, or cluster states around the chimera are studied in detail in previous studies~\cite{twocluster, prerequisite}}. However, finding the proper connection between thermodynamics and dynamics of coupled systems within a nonequilibrium environment is always elusive and crucial. Only very recently, chimera state in a continuum chemical system is characterized thermodynamically by leveraging the nonequilibrium thermodynamic framework of chemical reaction network (CRN)~\cite{Rao2016NonequilibriumThermodynamics, Falasco2018InformationPatterns} to reveal the guiding role of the information uncertainty principle in the evolution of chimera energetics~\cite{pkgg3}. Nevertheless, this thermodynamic study is confined solely to the chimera state only. Some recent developments have also addressed the thermodynamic cost of the coherent biochemical oscillations~\cite{TUfree, barato} within the stochastic thermodynamic framework.~{In this current  investigation, we aim to expand the scope of this line of research to encompass more general scenarios by incorporating diverse generic states derived around the chimera state within a coupled system and inspecting the dynamic and thermodynamic signatures of these generic collective states.} The inclusion of transitions between different dynamical states in this study, ranging from coherence and incoherence to chimeras and cluster states, can aid in establishing a dynamic and thermodynamic connection among these states, and thus broaden the understanding regarding the emergence, stability, generalization, and exclusiveness of those states and transitions within a coupled chemical system across various settings. In addition, the transitions between the coherent and incoherent states of collective systems are ubiquitous phenomena~\cite{mainFaraday, EBraun_1991, Granulargas}, and a general understanding of the transition between coherent and incoherent states remains a central issue in the study of coupled nonlinear systems. So a systematic and complete thermodynamic study of parametric transitions would shed light on the fundamental underlying fingerprints and nature of those similar transitions.	
	
	Here, we generate different spatiotemporal states in a generic continuum chemical oscillator system where we impose the effective coupling at the level of amplitude dynamics. Different spatiotemporal states are obtained by controlling the coupling parameters of the system, while the internal parameters of an individual chemical oscillator are kept at fixed values. The chemical oscillators are at the Hopf instability regimes, and hence a modified complex Ginzburg-Landau equation (MCGLE)~\cite{mcgle1st,mcgle2} is enacted to encapsulate the amplitude of the collective dynamics.~{To facilitate a connection between the dynamics of a globally coupled chemical system and its nonequilibrium thermodynamic depiction, we have here adopted a similar ansatz as in ref.~\cite{pkgg3}.} Our ansatz posits that the concentration fields of the spatiotemporal states can be acquired by combining the numerical solution of the MCGLE into the linear stability description of the reaction-diffusion system (RDS). After deriving different spatiotemporal patterns of the coupled system, especially by presenting a transition between coherence and incoherence patterns mediated via chimera or similar states, we intend to ask some pivotal questions from a thermodynamic viewpoint: How much energetic and entropic costs do we need to pay in sustaining those states? How are different collective dynamic states at distinct coupling parameters thermodynamically connected? How can we identify these transitions in the collective dynamics from thermodynamic characterizations? To answer these questions, we have systematically quantified the central entities of the nonequilibrium thermodynamics corresponding to those spatiotemporal states.~{ Moreover, we have uncovered some intriguing resemblances between essential dynamic and thermodynamic quantities, irrespective of the collective states and these similarities can hold significant relevance in the context of the connection between the dynamics and thermodynamics of coupled systems.}      
	
	The layout of the paper is as follows. First, we present the concentration dynamics and relevant amplitude equation of the globally coupled Brusselator system in Sec.~\ref{concentrationdynamics}. In the next section, elements of the nonequilibrium thermodynamics are properly formulated. Results and discussion are provided in Sec.~\ref{subsec7}. Finally, we conclude the work in Sec.~\ref{sec13}.         
	\section{\label{concentrationdynamics}CONCENTRATION DYNAMICS OF GLOBALLY COUPLED SYSTEM}         
	\subsection{\label{subsecI}Brusselator dynamics}
	Brusselator~\cite{Prigogine1968SymmetryII, Nicolis1977Self-organizationFluctuations}, a prototypical model for investigating oscillatory and cooperative behaviors in chemical and biological systems, is considered for demonstrating the transition between coherence and incoherence mediated via chimera or other similar states. The following chemical reactions describe the Brusselator model: 
	\begin{equation}
		\begin{aligned}
			\rho&=1:&\ch{A&<=>[\text{k\textsubscript{1}}][\text{k\textsubscript{-1}}] X}\\
			\rho&=2: &\ch{B + X&<=>[\text{k\textsubscript{2}}][\text{k\textsubscript{-2}}]Y + D}\\
			\rho&=3:& \ch{2 X + Y&<=>[\text{k\textsubscript{3}}][\text{k\textsubscript{-3}}]3X}\\
			\rho&=4:&\ch{X&<=>[\text{k\textsubscript{4}}][\text{k\textsubscript{-4}}]E}
		\end{aligned}
		\label{crn}
	\end{equation}
	with $'\rho'$ being the reaction step label. From this Brusselator chemical reaction network (CRN), we construct the stoichiometric matrix,
	\begin{gather}
		S_{\rho}^{\sigma}=
		\bordermatrix{ ~ & R_{1} & R_{2}&R_{3}&R_{4}\cr
			X&1 &-1&1&-1\cr
			Y&0&1&-1&0 \cr
			A&-1&0&0&0 \cr
			B&0&-1&0&0\cr
			D&0&1&0&0 \cr
			E&0&0&0&1\cr}.\label{st}
	\end{gather} 
	The species in this reaction network are separated into two disjoint sets: $\{X, Y\}\in I$ of intermediate species having dynamic concentration, and $\{A, B, D, E\}\in C$ of chemostatted species with a constant homogeneous concentration. The concentration dynamics of the species, {$\sigma$} of the CRN evolves according to a general rate equation,
	\begin{equation}
		\label{genform}
		\dot{z_{\sigma}}=\sum_{\rho}S_{\rho}^{\sigma}j_{\rho}+\delta_{\sigma}^{C}J_{C},
	\end{equation}
	with $z_{\sigma}$ being the concentration of the $\sigma$ species, $j_{\rho}$ being the net flux for the $\rho$ reaction step, $\delta_{\sigma}^{C}$ being the Kronecker delta and $J_{C}$ being the external flux related to the chemostatted species only. 
	
	The Brusselator CRN described above obeys the following rate equation, 
	\begin{widetext}
		\begin{equation}
			\begin{aligned}
				\dot{x}&={k_1}a-({k_2}b+k_4+k_{-1})x+(k_{-2}d+{k_3}x^2)y-k_{-3}x^3+k_{-4}e\\
				\dot{y}&={k_2}bx-k_{-2}dy+k_{-3}x^3-{k_3}x^2y 
			\end{aligned}
			\label{dynamic}	
		\end{equation}
	\end{widetext}
	with $x=[X]$, $y=[Y]$, $a=[A]$, $b=[B]$, $d=[D]$, $e=[E]$ denoting concentrations of species. Eq.~\eqref{dynamic} yields steady-state of the system as, $x_{0}=\frac{k_1a+k_{-4}e}{k_{-1}+k_4}$,  $y_{0}=\frac{({k_2}b+k_{-3}x_0^2)x_0}{{k_{-2}d}+{k_3}x_0^2}$. 
	\subsection{\label{subsecII}Hopf instability in the Brusselator}
	Now implementing the linear stability analysis around the steady state, $(x_{0},y_{0})$, the critical value of the control parameter for the onset of Hopf instability is derived as $b_{cH}=\frac{k_{4}}{k_{2}}+\frac{k_{1}^2 k_{3}}{k_{2}{k_4}^2}a^2$, under the assumption of a much higher forward reaction rate than the reverse ones, i.e., $k_{\rho}\gg k_{-\rho}$. The corresponding critical frequency of the oscillation is, $f_{cH}=\sqrt{\frac{k_{1}^2 k_{3}}{k_4}}a$ and critical eigenvector reads
	$U_{cH}=
	\begin{pmatrix}
		1+\frac{i}{a}\sqrt{\frac{k_4}{k_3}}\frac{1}{k_1}&,
		-(1+\frac{{k_4}^3}{k_3 {k_1}^2}\frac{1}{a^2}) 
	\end{pmatrix}^\textbf{T}$. In the presence of self-diffusion, the Brusselator dynamics in eq.~\eqref{dynamic} can be represented as a reaction-diffusion system (RDS) {in one spatial dimension $r\in [0,l]$ as,
		\begin{widetext}
			\begin{equation}
				\begin{aligned}
					\dot{x}&={k_1}a-({k_2}b+k_4+k_{-1})x+(k_{-2}d+{k_3}x^2)y-k_{-3}x^3+k_{-4}e+D_{11}x_{rr}\\
					\dot{y}&={k_2}bx-k_{-2}dy+k_{-3}x^3-{k_3}x^2y+D_{22}y_{rr}, 
					\label{ddynamic}
				\end{aligned}
			\end{equation}
		\end{widetext}
		where $D_{11}$, $D_{22}$ being the constant self-diffusion coefficients of intermediate species $X$ and $Y$, respectively.} The RDS of the Brusselator assumes the critical value of the control parameter $b$ for the Hopf instability as, $b_{ctw}=\frac{k_{4}}{k_{2}}+\frac{k_{1}^2 k_{3}}{k_{2}{k_4}^2}a^2+\frac{(D_{11}+D_{22})}{k_2}q^2$. $q$ is the wavenumber that obeys $q=\frac{2n\pi}{l}$ for periodic boundary conditions in the finite domain, $l$ with $n$ being an integer. However, to simplify our investigation, we have considered the wavenumber to be $q=0$. 
	\subsection{\label{subsecIII} Modified amplitude equation for globally coupled system}
	We can capture concentration dynamics of the nonlinear RDS by incorporating a complex entity, amplitude, with the standard linearized description of the system. The amplitude near the onset of the Hopf instability is derived from the CGLE. The normal form of the CGLE~\cite{Nicolis1995IntroductionScience, Cross2009PatternSystems} in a spatially extended system is 
	\begin{equation}
		\frac{\partial Z}{\partial t}=\lambda Z -(1-i\beta)\mid Z \mid ^2Z+(1+i \alpha)\partial_{r}^2 Z,
		\label{ncgle}
	\end{equation}
	with $Z$ being the complex amplitude field and $\lambda$, $\beta$, and $\alpha$ being coefficients containing the RDS parameters. Coefficients $\alpha$ and $\beta$ of this normal form of CGLE are determined by Krylov-Bogolyubov (KB) averaging method~\cite{krylov1949introduction, pkgg}. For the Brusselator model, the coefficients of the CGLE are
	$\alpha=\frac{\Omega(D_{22}-D_{11})}{(D_{11}+D_{22})}$, 
	$\beta=\frac{p_2}{p_1}\frac{1}{3a}$, and  $\lambda =\frac{b-1-a^2}{2}$ with the ratio of correction factors, $\frac{p_1}{p_2}$ being $\frac{4-7a^2+4a^4}{(2+a^2)}$. {The parameter $\Omega$ in the coefficient, $\alpha$ of CGLE is acquired from KB method as, $\Omega=a$, with $a$ being the concentration of chemostatted species, $A$.}   
	
	Now for the coupled continuum system with the global coupling at the level of the amplitude, the amplitude equation, CGLE in eq.~\eqref{ncgle} modifies into the MCGLE, 
	\begin{eqnarray}
		\frac{\partial Z}{\partial t}=\lambda Z -(1-i\beta)\mid Z \mid ^2Z+(1+i \alpha)\partial_{r}^2 Z \nonumber \\                     
		-(\lambda+i\nu) \left\langle Z \right\rangle
		+(1-i\beta)\left\langle \mid Z\mid ^2Z \right\rangle
		\label{gcncgle}
	\end{eqnarray}
	with $\left\langle...\right\rangle$ denoting the spatial average. This nonlinear global coupling scheme yields an oscillatory mean field when we take the spatial average over eq.~\eqref{gcncgle}. The oscillatory mean field is $\left\langle Z \right\rangle=Z_0=\eta \exp(-i\nu t)$ with $\eta$ and $\nu$ being the amplitude and the frequency of the oscillation, respectively. From eq.~\eqref{gcncgle}, we get the threshold value of the $\eta$ as $\eta_c=\sqrt{\frac{\lambda}{2}}$, below which uniform oscillation becomes unstable irrespective of other parameters. We can view $\eta$ as the coupling strength of the system. 
	\subsection{\label{subsecIV}Concentration fields of intermediate species}
	In this article, eq.~\eqref{gcncgle} is solved numerically by the pseudospectral method incorporated with an exponential time stepping algorithm\cite{COX2002}. We then combine the amplitude field obtained from the MCGLE simulation with the standard linearized description of the single nonlinear system to acquire the collective concentration dynamics of intermediate species, ${z_I}$ of the Brusselator system,
	\begin{equation}
		{z_I}_{H}={z_I}_{0}+Z_{M}U_{cH}\exp(i f_{cH}t)+C.C.
		\label{hwave}
	\end{equation} 
	Here $Z_{M}$ is the numerically acquired amplitude field from the eq.~\eqref{gcncgle}, and ${z_I}_{0}$ is steady-state
	values of two intermediate species. 
	\section{\label{thermoelements}ELEMENTS OF NONEQUILIBRIUM THERMODYNAMICS}
	Forces regarding reactions, diffusion, and the nature of chemostatted species distribution generate fluxes in the system. The presence of forces and fluxes enables energy and entropy flow within the nonequilibrium system. In this investigation, the solvent of the dilute solution is viewed as a special chemostatted species that plays the role of a thermal reservoir by keeping the system at isothermal and isobaric conditions.   	
	\subsection{Entropy production rate}
	In this study, fluxes of CRN follow the mass action law, $j_{\pm \rho}=k_{\pm\rho}\prod_{\sigma}z^{v_{\pm\rho}^{\sigma}}_{\sigma}$ with $'+'$ and $'-'$ label denoting the forward and backward reaction, respectively and ${v_{\pm\rho}^{\sigma}}$ is the number of the species $\sigma$. The force in the CRN is reaction affinity~\cite{Prigogine1954ChemicalDefay.}, $f_{\rho}=-\sum_{\sigma}{S_{\rho}^{\sigma}\mu_{\sigma}}$ with $S_{\rho}^{\sigma}=v_{{-}\rho}^{\sigma}-v_{{+}\rho}^{\sigma}$ being the stoichiometric coefficient of species and $\mu_{\sigma}=\mu_{\sigma}^o+\ln{\frac{z_{\sigma}}{z_0}}$ being the nonequilibrium chemical potential. Additionally,  $z_0$  is the solvent concentration, and $\mu_{\sigma}^o$ is standard-state chemical potential. The local detailed balance condition reads $\ln{\frac{k_{+\rho}}{k_{-\rho}}}=-\sum_{\sigma}{S_{\rho}^{\sigma}\mu_{\sigma}^o}$. Thus, the reaction affinities are quantified in terms of the reaction fluxes as $f_{\rho}= \ln{\frac{j_{+\rho}}{j_{-\rho}}}$. Eventually, utilizing flux-force form, the entropy production rate (EPR) along the whole chemical reaction pathways reads~\cite{Kondepudi2014ModernThermodynamics},
	$\frac{d\Sigma_{R}}{dt}=\frac{1}{T}\int  dr \sum_{\rho} (j_{+\rho}-j_{-\rho}) \ln{\frac{j_{+\rho}}{j_{-\rho}}}$ with $T$ being the constant absolute temperature set by solvent. Further, diffusion of the internal species can also contribute to the total entropy production rate, and it can be expressed as, 
	$\frac{d\Sigma_{D}}{dt}=\frac{1}{T}\int dr \Big[ D_{11}\frac{{\parallel{\frac{\partial x}{\partial r} }\parallel}^2}{x}+D_{22}\frac{{\parallel{\frac{\partial y}{\partial r} }\parallel}^2}{y}\Big] \label{eprdd}$. By integrating over temporal dimension instead, we easily get the spatial (local) entropy production corresponding to the reaction and diffusion, $\frac{\partial \Sigma_{R}}{\partial r}$ and $\frac{\partial \Sigma_{D}}{\partial r}$, respectively from the EPR expressions. This total entropy production is always positive by the second law of thermodynamics. 
	
	\subsection{Semigrand Gibbs free energy}
	\begin{figure}
		\centering	
		\includegraphics[width=\columnwidth]{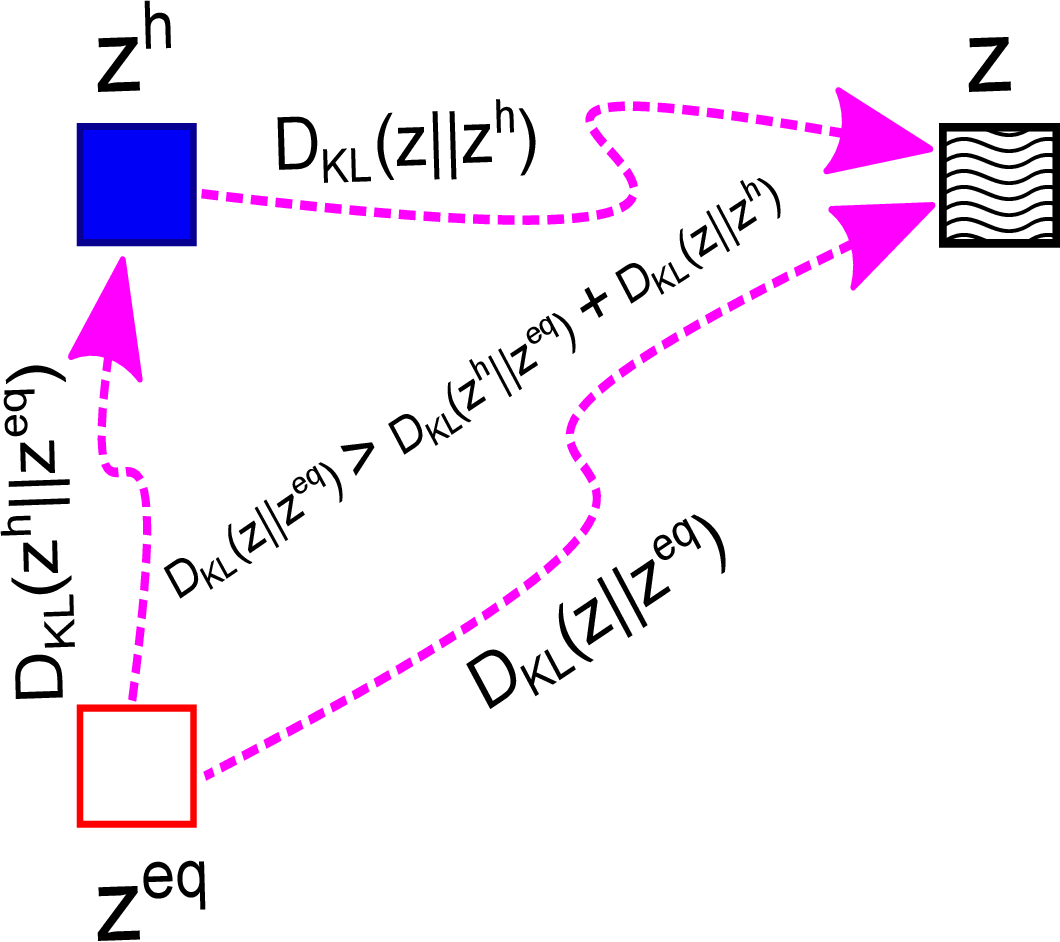}
		\caption{\label{illustration}~{Schematic illustration of potential transitions between three distinct states and their corresponding information-theoretic costs. Here, $z^{eq}$ represents the concentration of the system under detailed balance conditions, obtained from the reference chemical potential. $z^h$ corresponds to the nonequilibrium homogeneous state, while $z$ represents a nonequilibrium state with a specific pattern in the coupled system.Transitions between these different states are characterized using the Kullback-Leibler divergence, $D_{KL}$, of concentration distributions.}}
	\end{figure}
	For a proper energetic description of the nonequilibrium CRN, it is crucial to identify the conservation laws~\cite{Alberty2003ThermodynamicsReactions}. We obtain conservation law from the stoichiometric matrix $S_{\rho}^{\sigma}$ in eq.~\eqref{st} by acquiring left null  eigenvectors,  
	$\sum_{\sigma}{l_{\sigma}^{\lambda}S_{\rho}^{\sigma}}=0$, where $\{l_{\sigma}^{\lambda}\}\in \mathbb{R}^{(\sigma-w )\times \sigma}, w=rank(S_{\rho}^{\sigma})$. Conservation laws of the CRN further delineate globally conserved quantities of a closed system known as the components, $L_{\lambda}=\sum_{\sigma}{l_{\sigma}^{\lambda}}z_{\sigma}$. 
	
	When the closed system is opened by chemostating, conservation laws can be broken, and components corresponding to the broken conservation laws, $L_{\lambda_b}$, no longer remain the globally conserved quantities. We can disjoin the set of chemostatted species in the open system into two subsets $\{C\}=\{C_{b}\}\cup\{C_{u}\}$, based on its participation in breaking a conservation law. Labels $u$ and $b$ specify unbroken and broken ones, respectively. In this investigation, chemostatted species $A$ and $B$ are selected as elements of the set $C_{b}$, and the rest of the chemostatted species belong to the set $C_{u}$. 
	
	As the conservation laws are broken in the open system, nonequilibrium Gibbs free energy is not the proper entity to capture the energetics of the system. Instead, a transformed potential analogous to the grand potential of equilibrium thermodynamics has been employed in recent literature~\cite{Falasco2018InformationPatterns, Raoconservationlaw, Rao2016NonequilibriumThermodynamics}. To obtain this transformed potential, we first quantify the concentration of exchanged moieties corresponding to the broken components as $M_{C_{b}}=\sum_{\lambda_b}{l_{C_{b}}^{\lambda_b}}^{-1}L_{\lambda_b}$, with ${l_{C_{b}}^{\lambda_b}}^{-1}$ being derived from the inverse of the square and nonsingular matrix ${l_{C_{b}}^{\lambda_b}}.$ The energetics corresponding to the exchanged moieties is then captured by $\mu_{C_b}M_{C_{b}}$. Eventually, the proper energy content of the open CRN is acquired by {subtracting} the energetic cost of moieties exchange from the nonequilibrium Gibbs free energy $\mathcal{G}=G-\sum_{C_b}{\mu_{C_b}M_{C_b}},$ and $\mathcal{G}$ is recognized as the semigrand Gibbs free energy of the system~\cite{Falasco2018InformationPatterns, Raoconservationlaw}. The lower bound of semigrand Gibbs free energy is set at the reference equilibrium counterpart, $\mathcal{G}_{eq}$ acquired by exploiting the concentration fields $z_{\sigma}^{eq}$ which is derived from the reference chemical potential, $\mu_{\sigma}^{ref}$. $z_{\sigma}^{eq}$ represents the concentration of the system in a detailed balance condition {and is denoted as $z^{eq}$ in fig.~\ref{illustration}}. Most importantly, the semigrand Gibbs free energy and reference equilibrium counterpart are connected via an entity similar to the Kullback–Leibler (KL) divergence of information theory~\cite{Cover1999ElementsTheory} {as,
		$\mathcal{G}=\mathcal{G}_{eq}+D_{KL}(z_{\sigma} \vert \vert z_{\sigma}^{eq}),$ where the KL divergence or relative entropy, $D_{KL}(z_{\sigma} \vert \vert z_{\sigma}^{eq})$ can be expressed as $D_{KL}(z_{\sigma} \vert \vert z_{\sigma}^{eq})=z_{\sigma}\ln{\frac{z_{\sigma}}{z_{\sigma}^{eq}}}-(z_{\sigma}-{z_{\sigma}^{eq}})$ for the non-normalized concentration distribution~\cite{Falasco2018InformationPatterns}}.The KL divergence is always positive, {and it captures here the dissimilarity between the nonequilibrium concentration, $z_{\sigma}$ and reference equilibrium concentration, $z_{\sigma}^{eq}$}. 
	
	However, if our initial reference state is an arbitrary homogeneous state, $z_{\sigma}^{h}$ rather than an equilibrium state, {we can quantify the energetic cost of having the homogeneous state as 	$\mathcal{G}_h=\mathcal{G}_{eq}+D_{KL}(z_{\sigma}^h \vert \vert z_{\sigma}^{eq})$. Then, the difference between semigrand Gibbs free energy of nonequilibrium inhomogeneous and reference homogeneous states of the CRN is given by,    
		\begin{widetext}
			\begin{equation}
				\mathcal{G}-\mathcal{G}_h=D_{KL}(z_{\sigma} \vert \vert z_{\sigma}^{eq})-D_{KL}(z_{\sigma}^h \vert \vert z_{\sigma}^{eq})=D_{KL}(z_{\sigma}\vert \vert z_{\sigma}^{h})+\ln{\frac{k_{-1}x^h}{k_{+1}a}}(x-x^h)+\ln{\frac{k_{-1}k_{+3}y^h}{k_{+1}k_{-3}a}}(y-y^h),
				\label{maineq}
			\end{equation}
		\end{widetext}
		{where $D_{KL}(z_{\sigma}\vert \vert z_{\sigma}^{h})$ is the KL divergence between concentrations, $z_{\sigma}$ and $z_{\sigma}^{h}$. Notably, the above relation reveals that the information-theoretic cost (KL divergence) for switching from equilibrium to nonequilibrium inhomogeneous state via a nonequilibrium homogeneous state is lower than the direct transition between equilibrium and inhomogeneous states. The above analysis and discussion are practical and general, as applicable to arbitrary CRNs. These transitions and relations are illustrated in fig.~\ref{illustration}, where $\sigma$ is omitted in concentration representation to maintain concise notations.}}
	
	\subsection{Nonconservative work}
	In the general scenario, the dissipation in CRN is delineated in terms of the work and semigrand Gibbs free energy as
	$T\frac{d\Sigma_{D}}{dt}=\dot{w}_{driv}+\dot{w}_{ncon}-\frac{d \mathcal{G}}{dt}$, where $\dot{w}_{driv}$ quantifies work rate related to the external time-dependent manipulation of reference chemostats, and $\dot{w}_{ncon}$ represents the nonconservative work rate for sustaining steady currents of chemostatted species. Here for the autonomous system, $\dot{w}_{driv}$ vanishes. We also assume that chemostatted species have homogeneous distribution. Hence, there is no work due to the diffusion of chemostatted species, and the reference chemical potential $\mu_{C_b}^{ref}$ is $\mu_{C_b}$. This results in, $\mu_{a}^{ref}=\mu_{a}$ and $\mu_{b}^{ref}=\mu_{b}$ for the Brusselator system considered here. So fundamental forces corresponding to the reference chemostatted species are zero, i.e., $\mathcal{F}_{C_b}=0$. For the chemostatted species of the set $C_{u}$, $\mathcal{F}_{C_u}=\mu_{C_u}-\sum_{C_b}\mu_{C_b}\sum_{\lambda_b}{l_{C_b}^{\lambda_b}}^{-1}l_{C_u}^{\lambda_b}$ owing to the difference in chemical potentials of dissimilar chemostats~\cite{thermodynamicschemifcalwaves}. For the Brusselator system, this results in $\mathcal{F}_{d}=\mu_{d}-\mu_{b}$ and $\mathcal{F}_{e}=\mu_{e}-\mu_{a}$, implying $\mu_{d}^{ref}=\mu_{b}$, and $\mu_{e}^{ref}=\mu_{a}$. So the nonconservative work is only for chemostatted species $C_u$ and simplifies to $\dot{w}_{ncon}=-\sum_{C_u}(\mu_{C_u}-\mu_{C_u}^{ref})S_{\rho}^{C_u}j_{\rho}.$ This nonconservative work has been done to keep the concentration of the chemostatted species $C_u$ constant by offsetting the effect of chemical reactions.       
	
	\section{RESULTS AND DISCUSSION}\label{subsec7}
	\begin{figure*}[tb!]
		\includegraphics[width=\textwidth]{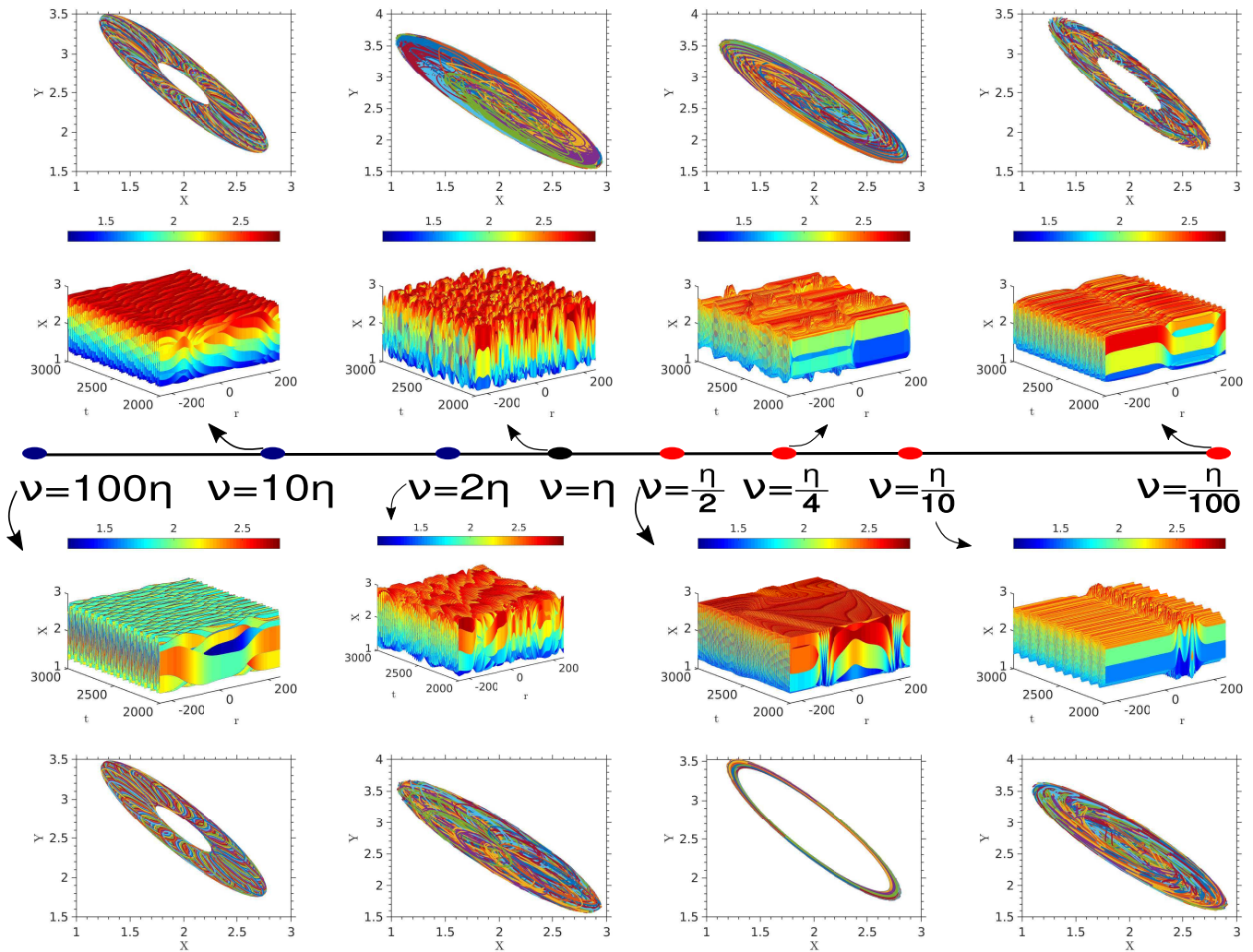}
		\caption{\label{chimeratransition}Transitions in the spatiotemporal concentration dynamics of the system due to the change in the mean-field (yielded in the nonlinear global coupling scheme) frequency keeping the amplitude of the mean-field fixed at $\eta=0.67\sqrt{\lambda}$. The magnitudes of the frequencies are indicated at the filled circle over the line. For different frequencies, the 3D concentration field of the activator is illustrated in the inner panel. {Here, color bars indicate the mapping of activator concentration values into the color map}. In the outer panel, X and Y spatiotemporal concentration dynamics are shown in phase portraits for the corresponding frequencies of the mean field. Other parameters are fixed at $D_{11}=4, D_{22}=3.2, l=500, a=2, b=5.24,\text{and}   k_{-{\rho}} =10^{-4}\ll k_{\rho} = 1$ for all illustrations.}
	\end{figure*}
	\begin{figure*}[tb!]
		\includegraphics[width=\textwidth]{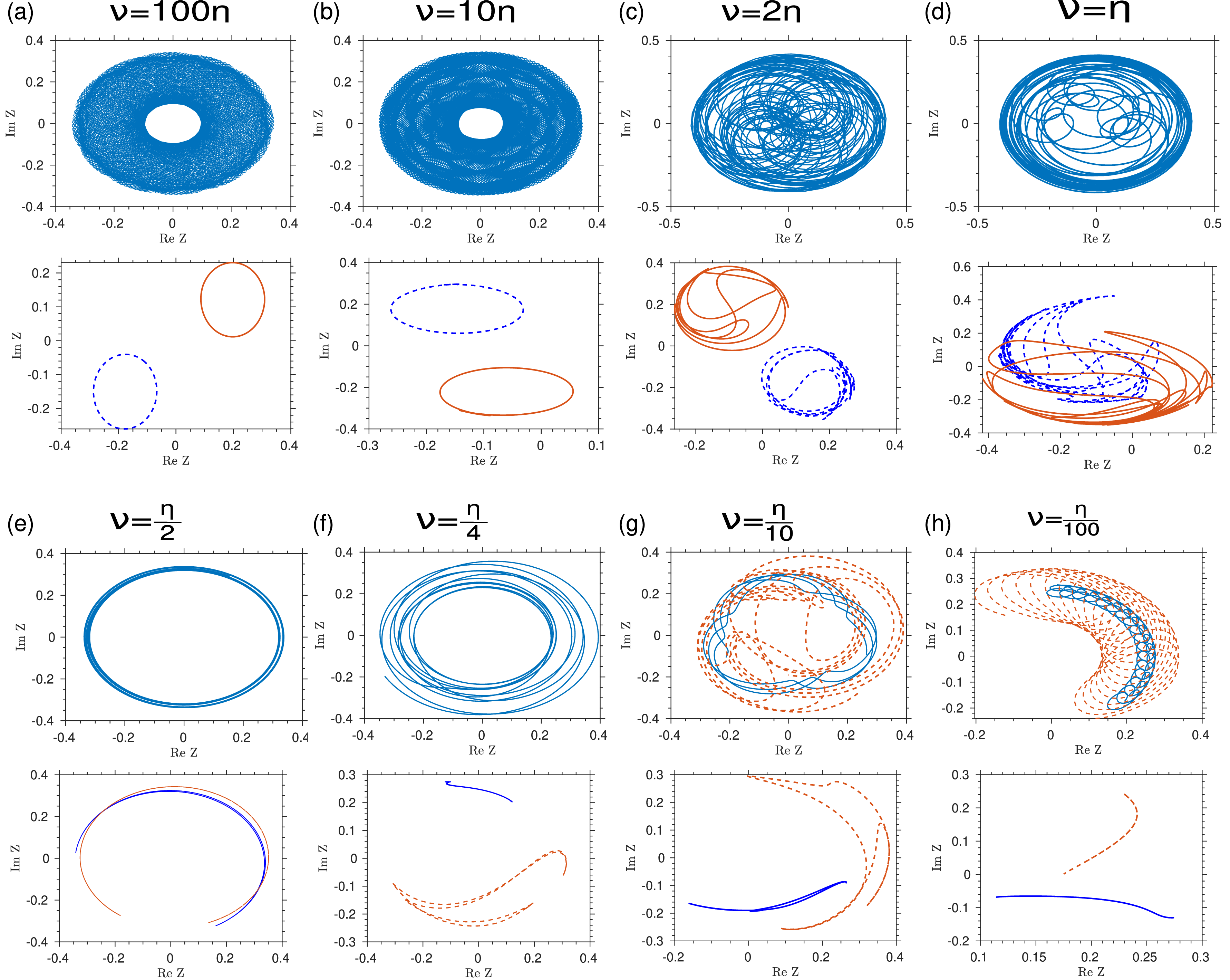}
		\caption{\label{complexplane}Amplitude dynamics corresponding to different states at the distinct coupling parameters are shown in the complex plane. Upper panel: Temporal dynamics of the amplitude at an arbitrary spatial node are demonstrated in a complex plane. For (g) and (h), we have shown temporal dynamics for two different spatial nodes (solid and dashed lines). Lower panel: {For each case of coupling parameters, the snapshot of the amplitude at an arbitrary time point is illustrated in the complex plane. Two profiles (solid and dashed lines) correspond to two different time points.}}
	\end{figure*}
	\begin{figure*}[tb!]
		\includegraphics[width=\textwidth]{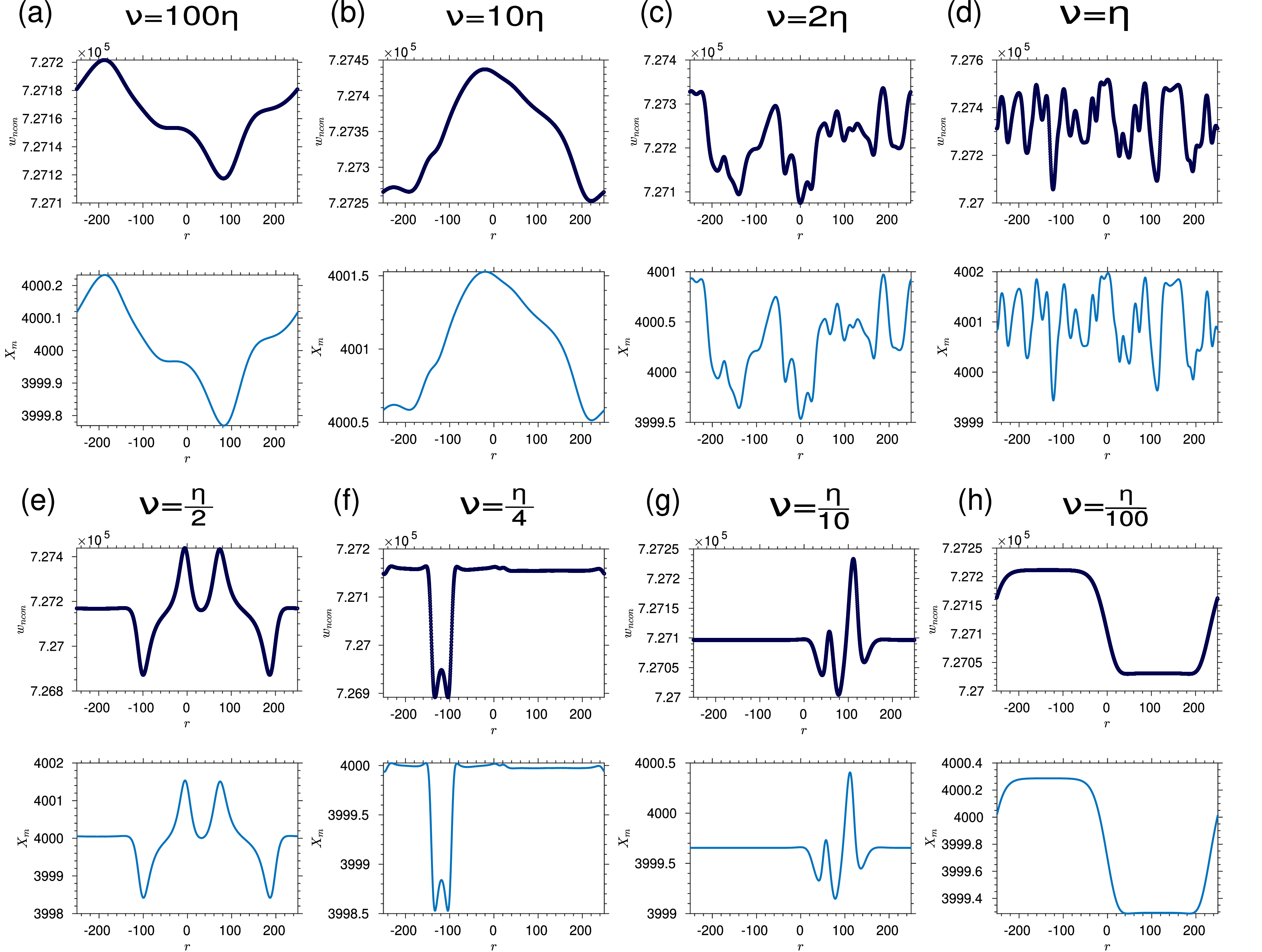}
		\caption{\label{work} Nonconservative work done on the system for different states related to distinct frequencies is shown in the upper panel. The corresponding time-integrated concentration profile, $X_m$, is illustrated in the lower panel. The nonconservative work and $X_m$ profiles are qualitatively analogous.}
	\end{figure*}	
	\begin{figure*}[tb!]
		\includegraphics[width=\textwidth]{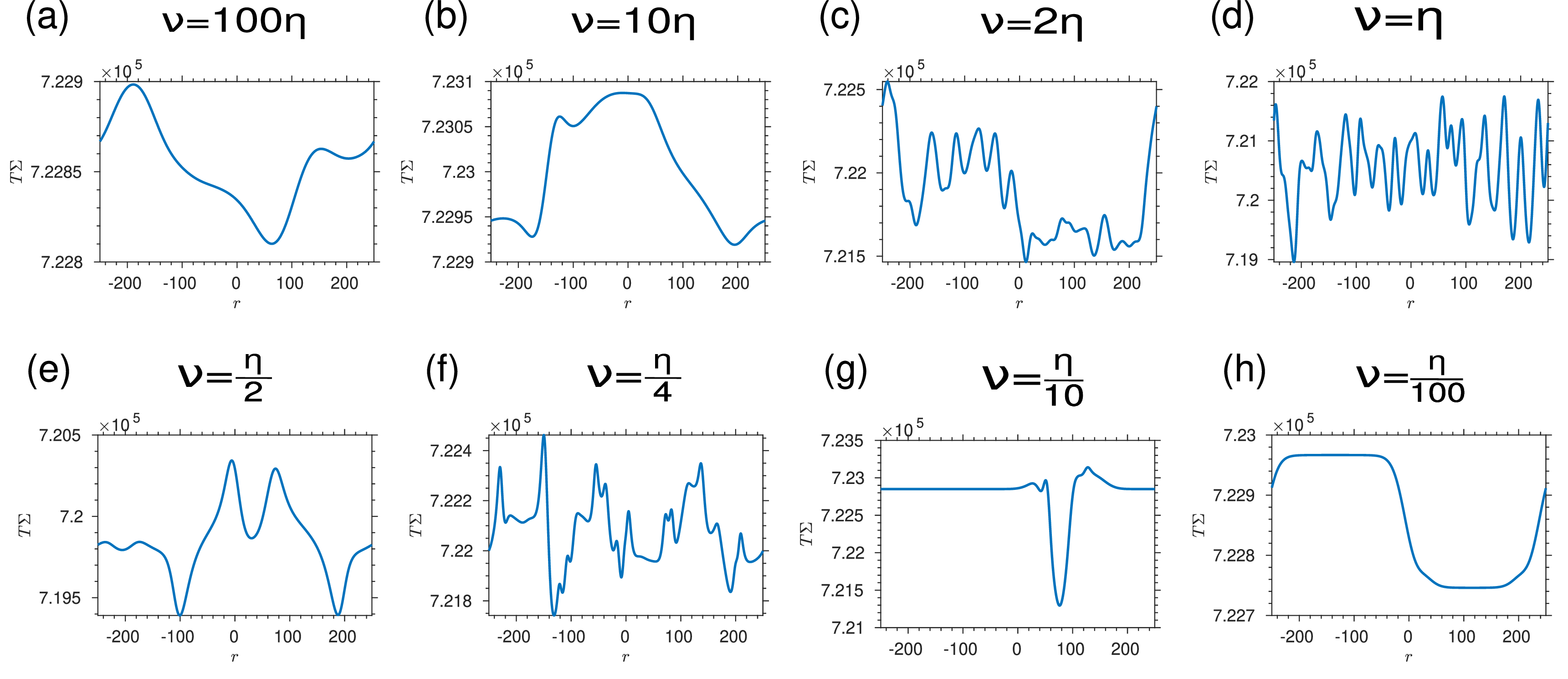}
		\caption{\label{epr}Entropy production of different states of the coupled system is presented over the spatial length of the system. (a), (b) and (h) correspond to coherent states at different frequencies. (c), and (e) illustrate the entropy production of two different transitioning states. (f), and (g) are entropy production for multichimera and chimera states, respectively. In (d), entropy production of the incoherent state is captured.}
	\end{figure*}
	\begin{figure*}[htb!]
		\includegraphics[width=\textwidth]{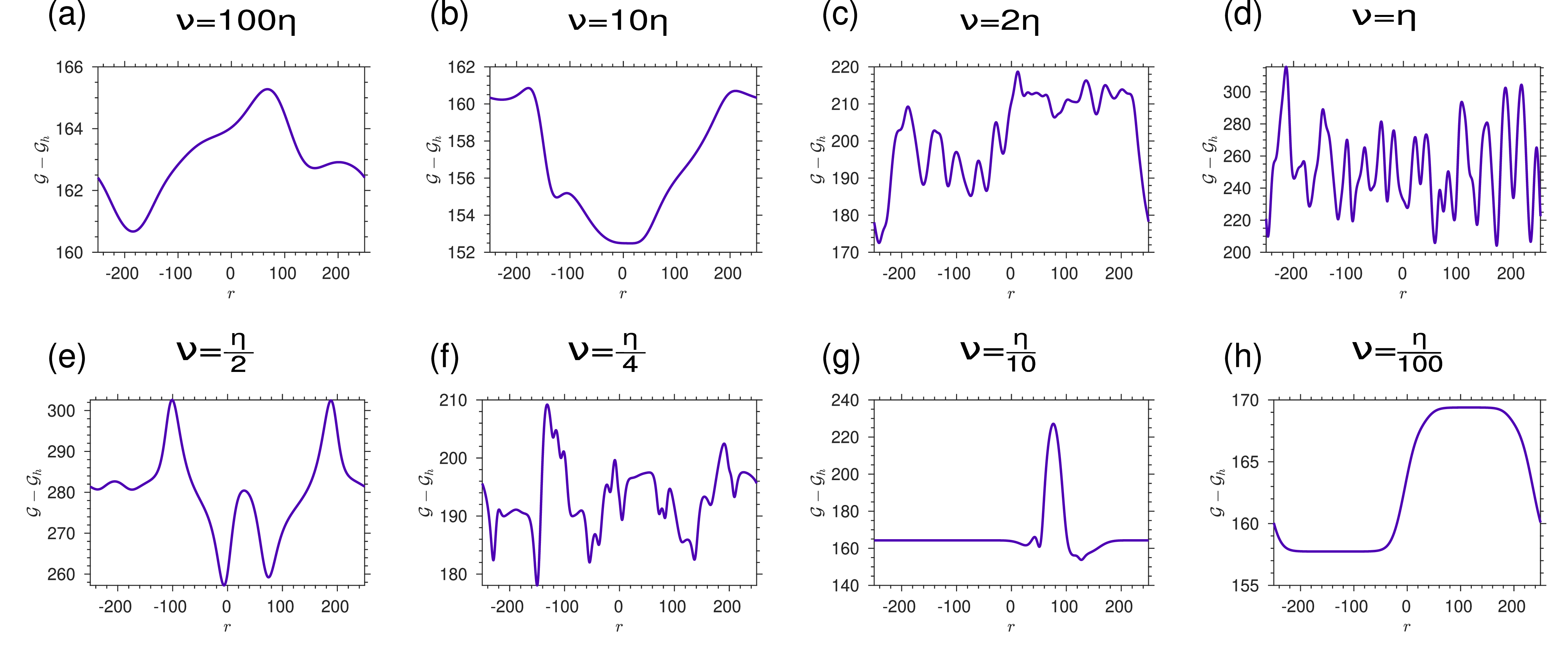}
		\caption{\label{Gibbs}Semigrand Gibbs free energy profiles corresponding to the states obtained via frequency variation are quantified. For all cases, the semigrand Gibbs free energy of a common initial homogeneous state is {subtracted} from the semigrand Gibbs free energy profiles of the states.}
	\end{figure*} 
	\begin{figure*}
		\centering	
		\includegraphics[width=\textwidth]{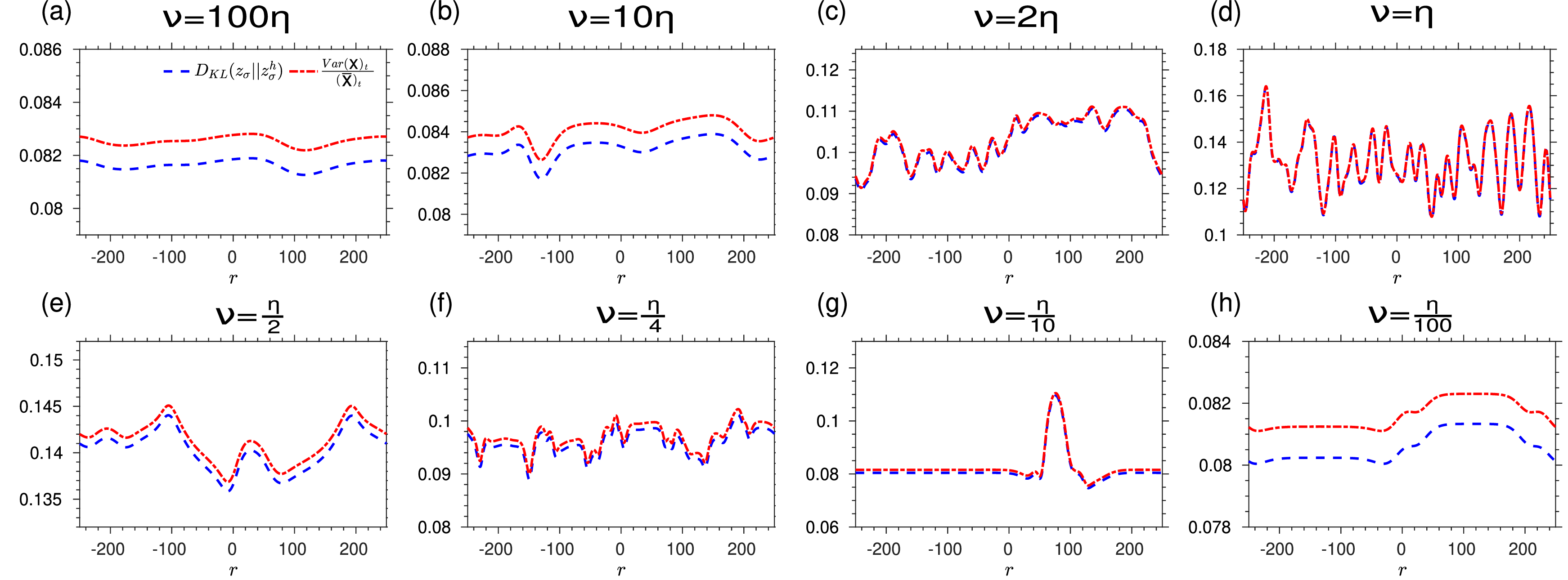}
		\caption{\label{relativevar} {The equivalence of Kullback-Leibler divergence between the nonequilibrium inhomogeneous and homogeneous concentrations with an entity resembling the Fano factor, denoted as $\frac{Var{(x)}_t}{(\overline{x})_t}$, where $Var{(x)}_t$ is the variance and $(\overline{x})_t$ is the average of the activator concentration in the inhomogeneous state over time, respectively. Here, $z_{\sigma}^h$ corresponds to the nonequilibrium homogeneous concentrations, while $z_\sigma$ represents concentrations of species at the nonequilibrium inhomogeneous state. The range for the two equivalence entities is very small in coherence states (a), (b), and (h), which highlights the seemingly pronounced difference between these entities.}}
	\end{figure*} 
	Different spatiotemporal states in the system are obtained by discretely varying the oscillatory mean-field (corresponding to the nonlinear global coupling) frequency, $\nu$ relative to the fixed amplitude $\eta=0.67\sqrt{\lambda}$ of the field. Hence, the frequency values are represented in terms of the mean-field amplitude in all illustrations. At every $\nu$ value, we initially keep the system at a uniform state. For all illustrations, the control parameter of the Brusselator is fixed at $b=5.24$, and other parameters of the RDS are $D_{11}=4$, $D_{22}=3.2$, $a=2$, \text{and}  $k_{-{\rho}} =10^{-4}\ll k_{\rho} = 1$. For the simulation of the MCGLE, a timestep of size $0.01$ is taken, and the system length $l=500$ is divided into 2048 grid points. Periodic boundary conditions are implemented. For every state, amplitude snapshots are taken between time $t=2000$ and $t=3000$. 
	
	In this investigation, we use quantitative metrics, the strength of incoherence (SI), $S$, and discontinuity measure (DM), $H$~\cite{laks} originally constructed for nonlocally coupled systems to characterize the different dynamical states. Similarly, we devise these statistical measures (see APPENDIX A) based on the local standard deviation for identifying dynamical states of the globally coupled continuum chemical oscillatory system. The combinations of $S$ and $H$ can aid in recognizing the dynamical states. More specifically, $(S=0, H=0)$, $(S=1, H=0)$, $(0<S<1, H=1)$, $(0<S<1, 2\leq H\leq\frac{P}{2})$ denote coherent, incoherent, chimera, and multichimera states, respectively. In this investigation, we demonstrate the following collective states at distinct frequency values of the mean-field relative to the amplitude.  
	
	(a)\textit{Quasiperiodic coherent state:} When the frequency is sufficiently larger than the amplitude, for example, $\nu=100\eta$, the system shows a spatially coherent behavior in~\cref{chimeratransition}. Snapshots of the concentrations reveal closed orbits in the complex plane in the lower panel of~\cref{complexplane} (a). However, the temporal dynamics of an arbitrarily chosen spatial point are quasiperiodic and have modulated amplitude. Therefore, the spatiotemporal behavior in the phase portrait of $\nu=100\eta$ in~\cref{chimeratransition} emerges as a circular motion on a torus. We recognize this state as a quasiperiodic coherent state. Then, for a relatively lower frequency value, $\nu=10\eta$, the system occupies a similar quasiperiodic coherent state with a complex torus attractor in the phase portrait in~\cref{chimeratransition}. However, this state is less coherent than the previous one, as a very small distortion appears on the circular orbit of the snapshot in~\cref{complexplane} (b). In both states, we obtain SI and DM as $S=0$ and $H=0$ as expected for the coherent states of the system.
	
	(b)\textit{Tree-like transition state:} As we explore low frequencies, we find that the cyclic orbit eventually gets split and is converted into a circular arc in the phase portrait of the snapshots. For further lower frequencies (shown for $\nu=2\eta$), a multifold trajectory emerges in the lower panel of~\cref{complexplane} (c). This trajectory implies the generation of incoherence in the spatial dimension. The temporal dynamics of the state in the upper panel exhibit a transition from intricate quasiperiodic motion to chaotic flow over time; hence, the torus attractor no longer exists in the corresponding spatiotemporal phase portrait in~\cref{chimeratransition}. In the spatiotemporal field of the concentration, tree-like patterns~\cite{zakharova2020chimera} appear locally over some spatial regimes, and this tree-like pattern spreads more as we lower the frequencies (see $\nu=2\eta$ in~\cref{chimeratransition}). The combination of SI and DM values indicates a multichimera state~\cite{multichimera} at $\nu=2\eta$ due to the loss of spatial coherence at multiple parts of the concentration field. However, we would recognize these partially synchronized states as ``tree-like transitioning states". A similar transitioning state is also identified in Couette flow~\cite{Couetteflow}. 
	
	(c)\textit{Incoherent state:} The complex tree-like state then evolves into spatiotemporal intermittency~\cite{chate1987transition, kaneko1985spatiotemporal} for a slightly lower frequency. As we keep decreasing the frequency discretely, irregular patches scattered over the spatiotemporal dimension of the concentration dynamics develop around the equal value of the amplitude and frequency (~\cref{chimeratransition}). For $\nu=\eta$, we now have a very complex trajectory in the phase portrait of snapshots, and the temporal dynamics in the phase portrait follow a peculiar attractor in~\cref{complexplane} (d). This state is identified as an incoherent state, and the overall spatiotemporal nature of the incoherent is exhibited in the corresponding phase portrait of the activator-inhibitor concentration in~\cref{chimeratransition}. The entities SI and DM, with values $S=1$ and $H=0$, also support the existence of the incoherent state. 
	
	(d)\textit{Transitioning cluster state:} As we further lower the frequency, keeping the amplitude, $\eta$ fixed, the incoherent nature is decreased as the size of the patches increases significantly, and we have simpler spatiotemporal intermittency on the surface of the concentration (not shown here). Due to further phase variation, a transition from this spatiotemporal intermittency to a tree-like perturbation~\cite{zakharova2020chimera} state can occur. The tree-like perturbation state for $\nu=\frac{\eta}{2}$ has been illustrated in~\cref{chimeratransition}. The spatiotemporal turbulence nature in the phase portrait is now converted into a thin cyclic orbit. So, unlike the incoherent regimes, the temporal dynamics phase portrait in~\cref{complexplane} (e) demonstrates multiple cyclic attractors attributed to multirythemicity. In the lower panel, snapshots of spatial grids in the complex plane reveal that the system possesses two clear circular arc attractors at the higher time, which indicates the existence of at least two groups. Although the combination of SI and DM values indicates a multichimera state at $\nu=\frac{\eta}{2}$ due to the loss of spatiotemporal coherence multiple times, we identify this as an amplitude cluster state, argued to be the prerequisite of having chimera in the globally coupled system~\cite{prerequisite}. 
	
	(e)\textit{Multichimera state:} Now for even lower frequency, for instance, $\nu=\frac{\eta}{4}$, multiple modulated amplitude clusters emerge in the phase portrait of the temporal dynamics of the spatial grid in~\cref{complexplane} (e). On the other hand, the snapshots of the spatial grids have folded double-layer trajectories. The cluster states in this coupling parameter move irregularly with time, leading to a change in the position of the coherence and incoherence regimes. Hence, multiple sharp incoherence clusters grow in the spatiotemporal concentration in~\cref{chimeratransition} and thus generate the appearance of chimera states with the coexistence of multiple coherent and incoherent clusters. A chimera state with a chaotic motion of coherent and incoherent regimes in time was previously described as a finite-size effect~\cite{chaoticspatiotemporal}. Here, in the continuum system, we identify this spatiotemporal pattern as a multicluster chimera, as also hinted by the combination of SI and DM values. 
	
	(f)\textit{chimera state:} For further lower frequency, i.e., $\nu=\frac{\eta}{10}$, a conventional amplitude-mediated chimera state emerges in the spatiotemporal concentration field in~\cref{chimeratransition}. The combination of SI and DM, $(0<S<1, H=1)$ also matches the identification of the chimera state. In the phase portrait,~\cref{complexplane} (g), the modulated amplitude dynamics are obtained for the spatial grid of the coherence regime, and coexisting incoherence part (dashed line) yields more intricate phase trajectories. Corresponding snapshots of spatial dynamics illustrate distorted $\rho$-shaped profiles~\cite{NakagawaKuramoto, sethiasen} that get modified with time. The nonlinear nature of global coupling causes complexity in the $\rho$-shaped profiles. 
	
	{(g)}\textit{chimera to coherent transition:} As the frequency of the mean-field decreases, a {type-II cluster~\cite{prerequisite}} pattern emerges with {a modulated amplitude of} temporal dynamics. Initially, this pattern appears quasiperiodic, but as the value of $\nu$ decreases further, it begins to split (~\cref{complexplane} (h)). At $\nu=\frac{\eta}{100}$, we have two out-of-phase coherence clusters in the concentration field (see fig.~\ref{chimeratransition}). Phases of the clusters alternate after some time. There is a miniature incoherent patch at the boundary of the clusters. The combination of SI and DM, $(S=0, H=0)$, suggests this pattern as a coherent state. 
	
	For all these states, the time-averaged description of a transformed variable, $\mathcal{X}_{i,t}$ (see APPENDIX A) reflects an odd symmetry in their profiles (not shown), and this implies an even symmetric center of mass for the activator concentration. Regimes belonging to the coherent states of the concentration field have $\mathcal{X}_{i,t}$ around the origin of the vertical axis. In contrast, the regimes corresponding to the incoherent states deviate from that origin.

	
	The nonequilibrium thermodynamic entities corresponding to all the collective states mentioned above are illustrated in~\cref{work},~\ref{epr}, and~\ref{Gibbs}. From~\cref{work}, one can notice that the nonconservative work profiles (upper panel) of all the states qualitatively follow the time-integrated profiles of the activator concentration, $X_m$ (lower panel). This feature is due to the activator concentration in the dominant forward reaction currents, $j_{+2}$ and $j_{+4}$, which are part of the nonconservative work expression. Most of the nonconservative work is dedicated to spatial entropy production in~\cref{epr}. Another nonequilibrium thermodynamic element, the semigrand Gibbs free energy profiles of the different nonequilibrium states of the system are illustrated in~\cref{Gibbs}. The semigrand Gibbs free energy of the initial homogeneous state is subtracted from the semigrand Gibbs free energy of all states. Hence, these profiles contain the amount of information generated along the spatial dimension to achieve these states from a homogeneous reference state. As all the conservation laws are broken here, the semigrand Gibbs free energy represented at the local level is the proper thermodynamic potential of the system~\cite{thermodynamicschemifcalwaves}. Since the transition between different states occurs with variation in the ratio of the amplitude and frequency of the mean field, the semigrand Gibbs free energy and entropy production profiles at the local level go through drastic modification and thus aid in distinguishing the dissimilarities of different states. In all these cases, semigrand Gibbs free energy and spatial entropy production profiles are similar but completely out of phase. The entropy production and local free energy flow due to the diffusion flux are negligible here. { In~\cref{relativevar}, we have demonstrated the KL divergence, $D_{KL}(z_{\sigma}\vert \vert z_{\sigma}^{h})$ (dashed line) obtained between the nonequilibrium inhomogeneous state, $z_\sigma$ and homogeneous reference state, $z_{\sigma}^h$ of the coupled system and a measure resembling the Fano factor, denoted as $\frac{Var{(x)}_t}{(\overline{x})_t}$ (dash-dotted line) for different values of $\nu$. Here $Var{(x)}_t$ is the variance and $(\overline{x})_t$ is the average of the activator concentration in the inhomogeneous state over time, respectively. Figure~\ref{relativevar} provides compelling evidence that the KL divergence, $D_{KL}(z_{\sigma}\vert \vert z_{\sigma}^{h})$ and the Fano factor-like entity exhibit qualitative and quantitative similarity with the same degree of variability. More importantly, the semigrand Gibbs free energy expression in eq.~\eqref{maineq} contains $D_{KL}(z_{\sigma}\vert \vert z_{\sigma}^{h})$, and thus the above equivalence indicates a strong association between the proper nonequilibrium thermodynamic potential and variance of the concentration dynamics of the coupled system.}  
	
	For the coherence state corresponding to $\nu=100 \eta$ in~\cref{chimeratransition}, all the dynamic and thermodynamic entities in~\cref{work},~\ref{epr}, and~\ref{Gibbs} (a) exhibit a sine waveform with harmonic distortion. In the case of $\nu=10 \eta$ coherence state, the sine wave profiles in all these entities are distorted by a different harmonic. We note that spatial entropy production profile in~\cref{epr} (b) and semigrand Gibbs energy profile in~\cref{Gibbs} (b) corresponding to the coherence state at $\nu=10 \eta$ has an additional cusp. This cusp possibly manifests the appearance of slight distortion in concentration dynamics shown in the complex plane in fig.~\ref{complexplane} (b). Whereas in the case of the coherent state of much lower frequency, i.e., $\nu=\frac{\eta}{100}$, we have clipped sine waveforms with roughly equal amplitude in both top and bottom for the dynamic and thermodynamic entities in~\cref{work},~\cref{epr}, and~\cref{Gibbs} (h). {$D_{KL}(z_{\sigma}\vert \vert z_{\sigma}^{h})$ and the Fano factor-like entity for the coherence states in~\cref{relativevar} (a), (b), and (h) have remarkably small ranges, making even the slightest magnitude differences between these two measures appear significantly prominent.}  
	
	In the case of the tree-like transition state between coherence and incoherence states corresponding to $\nu=2\eta$, the nonconservative work in~\cref{work} (c) yields a more composite waveform. The nonconservative work profile has multiple notches, and irregular oscillatory behavior with varying amplitude appears over the spatial dimension due to the occasional loss of spatial coherence in the presence of the tree-like pattern. The entropy production and semigrand Gibbs free energy profiles develop stronger modulation than the coherence states and have one part with a more regular multifold spatial period than the out-of-phase another region. The erratic part of the thermodynamic profiles indicates that loss of coherence due to the tree-like pattern is more dominant in that spatial regime. For the incoherence state corresponding to $\nu=\eta$, thermodynamic entities in~\cref{work},~\ref{epr}, and~\ref{Gibbs} (d) exhibit more breakup of oscillations and continuous aperiodic behaviors emerge over spatial length.{ Due to the comparatively large range of $D_{KL}(z_{\sigma}\vert \vert z_{\sigma}^{h})$ and the Fano factor-like entity for the tree-like transition and incoherence states in \cref{relativevar} (c), and (d), respectively, the two measures seem to have a notable degree of overlap.}  
	
	Then, as we obtain a transitioning cluster state from the incoherent state, the dynamics of the system become more regular and the aperiodic continuous wave profile translates into a symmetric profile for nonconservative work in~\cref{work} (e). This profile can be described as a superposition of two distinct behaviors. One part is flat, and then there is a transition to a double-peaked structure with a comparatively higher amplitude. In both the entropy production and semigrand Gibbs energy profiles, ~\cref{epr}, and~\ref{Gibbs} (e), two almost symmetric structures with the same phase but varying magnitudes are interconnected. At the point of connection between these structures, maxima appear in the semigrand Gibbs free energy profile. Unlike the nonconservative work profile, both the entropy production and semigrand Gibbs energy profiles display a wavy structure instead of a flat region. {Figure \cref{relativevar} (e) indicates that the range of the similar entities $D_{KL}(z_{\sigma}\vert \vert z_{\sigma}^{h})$ and $\frac{Var{(x)}_t}{(\overline{x})_t}$ for this transitioning cluster state falls within an intermediate level between that of the coherence and incoherence states.}   
	
	For the multichimera state, at $\nu=\frac{\eta}{4}$, the nonconservative work profile~\cref{work} (f) has a global minimum with spatial double-periodic structure. There are also some other secondary minima and maxima, which are not prominent markers. For the entropy production and semigrand Gibbs free energy, the profiles display completely distinct signatures comparing to the corresponding work profile, as can be seen from the~\cref{epr} and~\cref{Gibbs} (f), respectively. Entropy production and semigrand Gibbs free energy have several intricate structures with multiple bumps of different amplitudes corresponding to the incoherent regimes, and they are connected via small, relatively flat profiles corresponding to the coherence parts. This wave profile reflects that the multiple patches are not static over the spatiotemporal dimension; rather, the incoherence domains' boundaries have spatiotemporal erratic motion. {In this context, it is crucial to highlight that the nonconservative work is directly related to the addition of the KL divergence, $D_{KL}(z_\sigma \vert \vert z_\sigma^h)$ and entropy production for all the states. So despite the profound dissimilarity between the work profile and the individual profiles of entropy production and semigrand Gibbs free energy in this state, the former can indeed be derived from the scaled $D_{KL}(z_\sigma \vert \vert z_\sigma^h)$ and entropy production.} 
	
	For the conventional chimera state at $\nu=\frac{\eta}{10}$, the nonconservative work profile~\cref{work} (g) has a flat line corresponding to the coherence state, and an asymmetric profile belongs to the incoherence regime. For entropy production over spatial dimension in~\cref{epr} (g) exhibit hump structures with a global minimum between them. The global minima correspond to the center regime of the incoherence part of the chimera. The hump in this entropy production profile is a marker for the transition from coherence to incoherence. The semigrand Gibbs free energy in~\cref{Gibbs} (g) resembles a modulated pulse structure for the incoherent regime with a peak identifying the core of the incoherent state. Additionally, notches on both sides mark the transition from coherence to incoherence and correspond to the previously mentioned humps in the spatial entropy production profile. {The variation of entities $D_{KL}(z_{\sigma}\vert \vert z_{\sigma}^{h})$ and $\frac{Var{(x)}_t}{(\overline{x})_t}$ is also similar for the multichimera and chimera states in~\cref{relativevar} (f) and (g), respectively, and the range of these two entities is comparable to the range related to transitioning states.}  
	
	For the coherence, incoherence, and transitioning cluster state between incoherence and chimera, we get closely similar signatures for the nonconservative work and entropy production. However, we have distinct signatures for nonconservative work and entropy production profiles for the chimera, multichimera, and the tree-like transition state between the coherence and incoherence. As nonconservative work qualitatively reflects the time-integrated concentration dynamics, we also assert that system dynamics signatures at the level of the first raw moment differ from the spatial entropy production for the chimera, multichimera, and the tree-like transition state of the collective dynamics of the system.                             	
	
	\section{CONCLUSION}\label{sec13} To sum up, our work has identified the emergence of the different states and the transition between them in a globally coupled continuum chemical oscillatory system and quantified the corresponding nonequilibrium thermodynamic entities to capture the nonequilibrium thermodynamic signatures. In this regard, detailed comparative studies of these states are carried out, and thus acquired dynamic and thermodynamic signatures can be utilized to differentiate different dynamical states of chemical reaction networks qualitatively. The profiles of spatial entropy production in this study can be key diagnostic elements to detect the nature of transitions in diverse collective systems with different coupling schemes. The profiles of nonconservative work and semigrand Gibbs free energy for all the states are acquired here to reveal the connection between thermodynamic and dynamic entities and also among the thermodynamic entities.~{Notably, a concrete and intriguing connection between the semigrand Gibbs free energy and the variance of the activator concentration is asserted by displaying similarities between information-theoretic cost and Fano-factor-like measure. This result hold promise for advancing our understanding of complex systems in the fascinating intersection of thermodynamics and dynamics. Further, we observe that the Fano-factor-like measure for all the states is significantly lower than the constant value (=1) of the entity corresponding to a situation related to a Poisson process, which confirms that the underlying dynamics of these "under-dispersed" states are more complex and intricate than the Poisson process.}   
	
	Although we have obtained collective dynamics in a specific framework of the chemical oscillator, the principal constituent exploited for this nonequilibrium thermodynamic investigation of collective dynamics is common in many frameworks where such dynamical states were previously reported, and hence investigating similar thermodynamic entities is also viable for those systems. For instance, the thermodynamic characterization of different dynamical states in continuum settings could be extended to a ring of coupled oscillators, thus facilitating the investigation to reveal the connection between the chimera and other states in Kuramoto-type networks~\cite{Kotwal}. Following the recipe designated in this investigation, the nonequilibrium thermodynamic description of glycolytic oscillation~\cite{pkgg2} can be extended to collective dynamical states of such biological phenomena. The semigrand Gibbs free energy quantification of the different dynamic states and its information-theoretic connection conferred here can be crucial for elucidating different patterns of brain dynamics and function~\cite{kanikabrain}.~{Additionally, the valuable connection between the semigrand Gibbs free energy and variance of the activator concentration illustrated here for the global coupling framework of the chemical oscillator should be extended for local and nonlocal coupling schemes as well to assess the generality of this connection.}           
	
	\section{APPENDIX A: Classification of dynamic states using the strength of incoherence and discontinuity measure}
	\label{appen}
	Initially, from the spatiotemporal concentration field of the activator, $x$, we define a transformed variable over the whole time range $t$,  $\mathcal{X}_{i,t}=x_{i,t}-x_{i+1,t}$ with $i=1,2,..., N-1$ being the grid point over the spatial length, $l$. Then to differentiate the coherent state from the incoherent and chimera state, we calculate the standard deviation of the variable and take the time average, $\sigma=\left\langle\sqrt{\frac{1}{N}\sum_{i=1}^N[\mathcal{X}_{i,t}-<\mathcal{X}_{i,t}>]^2}\right\rangle_t$    
	with $<\mathcal{X}_{i,t}>=\frac{1}{N}\sum_{i=1}^N\mathcal{X}_{i,t}$ and $\left\langle...\right\rangle_t$ denoting the average over time. A vanishingly small $\sigma$ value indicates a coherent state, whereas relatively higher values of $\sigma$ imply incoherent and chimera states. However, as incoherent and chimera can have the same order of $\sigma$, we need another entity to distinguish chimera from incoherent states. So we define local standard deviation for a properly chosen bin number P, $\sigma_{loc}=\left\langle\sqrt{\frac{1}{n}\sum_{j=np-n+1}^{pn}[\mathcal{X}_{j,t}-<\mathcal{X}_{i,t}>]^2}\right\rangle_t$ with $n=\frac{N-1}{P}$ and $p=1, 2,..., P$. Now, we use a Heaviside step function, $\Theta(\delta-\sigma_{loc})$ with threshold, $\delta=0.0003|\mathcal{X}_{max}-\mathcal{X}_{min}|$ to distinguish these states. Finally, the strength of incoherence (SI) is defined as $S=1-\frac{\sum_{p=1}^{P}\Theta(\delta-\sigma_{loc}(p))}{P}.$ Additionally, the entity that aids in differentiating the multichimera from the chimera state is the discontinuity measure, $H=\frac{\sum_{i=1}^{P}|\Theta_i-\Theta_{i+1}|}{2}$.
	
	\section{APPENDIX B: Conservation laws of the Brusselator}
	\label{appen2}
	For the Brusselator reaction network, the conservation laws of the closed system comprise two linearly independent $(1\times 6)$ vectors,
	\begin{gather}
		l_{\sigma}^{\lambda=1}=  
		\bordermatrix{~&X&Y&A&B&D&E\cr
			&1&1&1&0&0&1\cr}\label{ce1}\nonumber
	\end{gather}
	and
	\begin{gather}  
		l _{\sigma}^{\lambda=2}=   
		\bordermatrix{~&X&Y&A&B&D&E\cr
			&0&0&0&1&1&0\cr}.\label{ce2}\nonumber
	\end{gather}
	Hence, components of the system are $L_1=x+y+a+e$ and $L_2=b+d$. From the chemostatted species of the chemical reaction network, any of the combinations $(A, B), (A, D), (E, B),$ and $(E, D)$ would break both the conservation laws in Brusselator. In this investigation, we select the combination, $(A, B)$ as the reference chemostatted species and the entries of the inverse of the identity matrix 
	$\begin{pmatrix}
		1 & 0 \\
		0 & 1  \\ 
	\end{pmatrix}$ are denoted by ${l_{C_{b}}^{\lambda_b}}^{-1}.$ Hence moieties concentrations are given here by $M_{A}=x+y+a+e$, and $M_{B}=b+d$. The reference chemical potential of intermediate species of the Brusselator is given by,    	 
	$\mu_{x}^{ref}=\sum_{C_b}\mu_{C_b}\sum_{\lambda_b}{l_{C_b}^{\lambda_b}}^{-1}l_{x}^{\lambda_b}=\mu_a$, and $\mu_{y}^{ref}=\sum_{C_b}\mu_{C_b}\sum_{\lambda_b}{l_{C_b}^{\lambda_b}}^{-1}l_{y}^{\lambda_b}=\mu_a$.
	
	\bibliography{sn-article}

\begin{thebibliography}{61}%
\makeatletter
\providecommand \@ifxundefined [1]{%
 \@ifx{#1\undefined}
}%
\providecommand \@ifnum [1]{%
 \ifnum #1\expandafter \@firstoftwo
 \else \expandafter \@secondoftwo
 \fi
}%
\providecommand \@ifx [1]{%
 \ifx #1\expandafter \@firstoftwo
 \else \expandafter \@secondoftwo
 \fi
}%
\providecommand \natexlab [1]{#1}%
\providecommand \enquote  [1]{``#1''}%
\providecommand \bibnamefont  [1]{#1}%
\providecommand \bibfnamefont [1]{#1}%
\providecommand \citenamefont [1]{#1}%
\providecommand \href@noop [0]{\@secondoftwo}%
\providecommand \href [0]{\begingroup \@sanitize@url \@href}%
\providecommand \@href[1]{\@@startlink{#1}\@@href}%
\providecommand \@@href[1]{\endgroup#1\@@endlink}%
\providecommand \@sanitize@url [0]{\catcode `\\12\catcode `\$12\catcode
  `\&12\catcode `\#12\catcode `\^12\catcode `\_12\catcode `\%12\relax}%
\providecommand \@@startlink[1]{}%
\providecommand \@@endlink[0]{}%
\providecommand \url  [0]{\begingroup\@sanitize@url \@url }%
\providecommand \@url [1]{\endgroup\@href {#1}{\urlprefix }}%
\providecommand \urlprefix  [0]{URL }%
\providecommand \Eprint [0]{\href }%
\providecommand \doibase [0]{https://doi.org/}%
\providecommand \selectlanguage [0]{\@gobble}%
\providecommand \bibinfo  [0]{\@secondoftwo}%
\providecommand \bibfield  [0]{\@secondoftwo}%
\providecommand \translation [1]{[#1]}%
\providecommand \BibitemOpen [0]{}%
\providecommand \bibitemStop [0]{}%
\providecommand \bibitemNoStop [0]{.\EOS\space}%
\providecommand \EOS [0]{\spacefactor3000\relax}%
\providecommand \BibitemShut  [1]{\csname bibitem#1\endcsname}%
\let\auto@bib@innerbib\@empty
\bibitem [{\citenamefont {Kuramoto}\ and\ \citenamefont
  {Battogtokh}(2002)}]{kuramoto2002coexistence}%
  \BibitemOpen
  \bibfield  {author} {\bibinfo {author} {\bibfnamefont {Y.}~\bibnamefont
  {Kuramoto}}\ and\ \bibinfo {author} {\bibfnamefont {D.}~\bibnamefont
  {Battogtokh}},\ }\bibfield  {title} {\bibinfo {title} {Coexistence of
  coherence and incoherence in nonlocally coupled phase oscillators},\
  }\href@noop {} {\bibfield  {journal} {\bibinfo  {journal} {Nonlinear Phenom.
  Complex Syst.}\ }\textbf {\bibinfo {volume} {5(4)}},\ \bibinfo {pages} {380}
  (\bibinfo {year} {2002})}\BibitemShut {NoStop}%
\bibitem [{\citenamefont {Abrams}\ and\ \citenamefont
  {Strogatz}(2004)}]{strogatz}%
  \BibitemOpen
  \bibfield  {author} {\bibinfo {author} {\bibfnamefont {D.~M.}\ \bibnamefont
  {Abrams}}\ and\ \bibinfo {author} {\bibfnamefont {S.~H.}\ \bibnamefont
  {Strogatz}},\ }\bibfield  {title} {\bibinfo {title} {Chimera states for
  coupled oscillators},\ }\href {https://doi.org/10.1103/PhysRevLett.93.174102}
  {\bibfield  {journal} {\bibinfo  {journal} {Phys. Rev. Lett.}\ }\textbf
  {\bibinfo {volume} {93}},\ \bibinfo {pages} {174102} (\bibinfo {year}
  {2004})}\BibitemShut {NoStop}%
\bibitem [{\citenamefont {Sethia}\ \emph {et~al.}(2008)\citenamefont {Sethia},
  \citenamefont {Sen},\ and\ \citenamefont {Atay}}]{sethiasen0}%
  \BibitemOpen
  \bibfield  {author} {\bibinfo {author} {\bibfnamefont {G.~C.}\ \bibnamefont
  {Sethia}}, \bibinfo {author} {\bibfnamefont {A.}~\bibnamefont {Sen}},\ and\
  \bibinfo {author} {\bibfnamefont {F.~M.}\ \bibnamefont {Atay}},\ }\bibfield
  {title} {\bibinfo {title} {Clustered chimera states in delay-coupled
  oscillator systems},\ }\href
  {https://link.aps.org/doi/10.1103/PhysRevLett.100.144102} {\bibfield
  {journal} {\bibinfo  {journal} {Phys. Rev. Lett.}\ }\textbf {\bibinfo
  {volume} {100}},\ \bibinfo {pages} {144102} (\bibinfo {year}
  {2008})}\BibitemShut {NoStop}%
\bibitem [{\citenamefont {Martens}\ \emph {et~al.}(2010)\citenamefont
  {Martens}, \citenamefont {Laing},\ and\ \citenamefont
  {Strogatz}}]{spiralchimera}%
  \BibitemOpen
  \bibfield  {author} {\bibinfo {author} {\bibfnamefont {E.~A.}\ \bibnamefont
  {Martens}}, \bibinfo {author} {\bibfnamefont {C.~R.}\ \bibnamefont {Laing}},\
  and\ \bibinfo {author} {\bibfnamefont {S.~H.}\ \bibnamefont {Strogatz}},\
  }\bibfield  {title} {\bibinfo {title} {Solvable model of spiral wave
  chimeras},\ }\href {https://link.aps.org/doi/10.1103/PhysRevLett.104.044101}
  {\bibfield  {journal} {\bibinfo  {journal} {Phys. Rev. Lett.}\ }\textbf
  {\bibinfo {volume} {104}},\ \bibinfo {pages} {044101} (\bibinfo {year}
  {2010})}\BibitemShut {NoStop}%
\bibitem [{\citenamefont {Motter}(2010)}]{motter2010spontaneous}%
  \BibitemOpen
  \bibfield  {author} {\bibinfo {author} {\bibfnamefont {A.~E.}\ \bibnamefont
  {Motter}},\ }\bibfield  {title} {\bibinfo {title} {Spontaneous synchrony
  breaking},\ }\href@noop {} {\bibfield  {journal} {\bibinfo  {journal} {Nature
  Physics}\ }\textbf {\bibinfo {volume} {6}},\ \bibinfo {pages} {164} (\bibinfo
  {year} {2010})}\BibitemShut {NoStop}%
\bibitem [{\citenamefont {Bordyugov}\ \emph {et~al.}(2010)\citenamefont
  {Bordyugov}, \citenamefont {Pikovsky},\ and\ \citenamefont
  {Rosenblum}}]{turbulent}%
  \BibitemOpen
  \bibfield  {author} {\bibinfo {author} {\bibfnamefont {G.}~\bibnamefont
  {Bordyugov}}, \bibinfo {author} {\bibfnamefont {A.}~\bibnamefont
  {Pikovsky}},\ and\ \bibinfo {author} {\bibfnamefont {M.}~\bibnamefont
  {Rosenblum}},\ }\bibfield  {title} {\bibinfo {title} {Self-emerging and
  turbulent chimeras in oscillator chains},\ }\href
  {https://link.aps.org/doi/10.1103/PhysRevE.82.035205} {\bibfield  {journal}
  {\bibinfo  {journal} {Phys. Rev. E}\ }\textbf {\bibinfo {volume} {82}},\
  \bibinfo {pages} {035205} (\bibinfo {year} {2010})}\BibitemShut {NoStop}%
\bibitem [{\citenamefont {Laing}(2015)}]{Lainglocal}%
  \BibitemOpen
  \bibfield  {author} {\bibinfo {author} {\bibfnamefont {C.~R.}\ \bibnamefont
  {Laing}},\ }\bibfield  {title} {\bibinfo {title} {Chimeras in networks with
  purely local coupling},\ }\href
  {https://link.aps.org/doi/10.1103/PhysRevE.92.050904} {\bibfield  {journal}
  {\bibinfo  {journal} {Phys. Rev. E}\ }\textbf {\bibinfo {volume} {92}},\
  \bibinfo {pages} {050904} (\bibinfo {year} {2015})}\BibitemShut {NoStop}%
\bibitem [{\citenamefont {Bastidas}\ \emph {et~al.}(2015)\citenamefont
  {Bastidas}, \citenamefont {Omelchenko}, \citenamefont {Zakharova},
  \citenamefont {Sch\"oll},\ and\ \citenamefont {Brandes}}]{quantumchimera}%
  \BibitemOpen
  \bibfield  {author} {\bibinfo {author} {\bibfnamefont {V.~M.}\ \bibnamefont
  {Bastidas}}, \bibinfo {author} {\bibfnamefont {I.}~\bibnamefont
  {Omelchenko}}, \bibinfo {author} {\bibfnamefont {A.}~\bibnamefont
  {Zakharova}}, \bibinfo {author} {\bibfnamefont {E.}~\bibnamefont
  {Sch\"oll}},\ and\ \bibinfo {author} {\bibfnamefont {T.}~\bibnamefont
  {Brandes}},\ }\bibfield  {title} {\bibinfo {title} {Quantum signatures of
  chimera states},\ }\href
  {https://link.aps.org/doi/10.1103/PhysRevE.92.062924} {\bibfield  {journal}
  {\bibinfo  {journal} {Phys. Rev. E}\ }\textbf {\bibinfo {volume} {92}},\
  \bibinfo {pages} {062924} (\bibinfo {year} {2015})}\BibitemShut {NoStop}%
\bibitem [{\citenamefont {Kruk}\ \emph {et~al.}(2018)\citenamefont {Kruk},
  \citenamefont {Maistrenko},\ and\ \citenamefont {Koeppl}}]{Selfpropelled}%
  \BibitemOpen
  \bibfield  {author} {\bibinfo {author} {\bibfnamefont {N.}~\bibnamefont
  {Kruk}}, \bibinfo {author} {\bibfnamefont {Y.}~\bibnamefont {Maistrenko}},\
  and\ \bibinfo {author} {\bibfnamefont {H.}~\bibnamefont {Koeppl}},\
  }\bibfield  {title} {\bibinfo {title} {Self-propelled chimeras},\ }\href
  {https://link.aps.org/doi/10.1103/PhysRevE.98.032219} {\bibfield  {journal}
  {\bibinfo  {journal} {Phys. Rev. E}\ }\textbf {\bibinfo {volume} {98}},\
  \bibinfo {pages} {032219} (\bibinfo {year} {2018})}\BibitemShut {NoStop}%
\bibitem [{\citenamefont {Hagerstrom}\ \emph {et~al.}(2012)\citenamefont
  {Hagerstrom}, \citenamefont {Murphy}, \citenamefont {Roy}, \citenamefont
  {H{\"o}vel}, \citenamefont {Omelchenko},\ and\ \citenamefont
  {Sch{\"o}ll}}]{hagerstrom2012experimental}%
  \BibitemOpen
  \bibfield  {author} {\bibinfo {author} {\bibfnamefont {A.~M.}\ \bibnamefont
  {Hagerstrom}}, \bibinfo {author} {\bibfnamefont {T.~E.}\ \bibnamefont
  {Murphy}}, \bibinfo {author} {\bibfnamefont {R.}~\bibnamefont {Roy}},
  \bibinfo {author} {\bibfnamefont {P.}~\bibnamefont {H{\"o}vel}}, \bibinfo
  {author} {\bibfnamefont {I.}~\bibnamefont {Omelchenko}},\ and\ \bibinfo
  {author} {\bibfnamefont {E.}~\bibnamefont {Sch{\"o}ll}},\ }\bibfield  {title}
  {\bibinfo {title} {Experimental observation of chimeras in coupled-map
  lattices},\ }\href@noop {} {\bibfield  {journal} {\bibinfo  {journal} {Nature
  Physics}\ }\textbf {\bibinfo {volume} {8}},\ \bibinfo {pages} {658} (\bibinfo
  {year} {2012})}\BibitemShut {NoStop}%
\bibitem [{\citenamefont {{Tinsley}}\ \emph {et~al.}(2012)\citenamefont
  {{Tinsley}}, \citenamefont {{Nkomo}},\ and\ \citenamefont
  {{Showalter}}}]{Tinsleynature2012}%
  \BibitemOpen
  \bibfield  {author} {\bibinfo {author} {\bibfnamefont {M.~R.}\ \bibnamefont
  {{Tinsley}}}, \bibinfo {author} {\bibfnamefont {S.}~\bibnamefont {{Nkomo}}},\
  and\ \bibinfo {author} {\bibfnamefont {K.}~\bibnamefont {{Showalter}}},\
  }\bibfield  {title} {\bibinfo {title} {{Chimera and phase-cluster states in
  populations of coupled chemical oscillators}},\ }\href
  {https://www.nature.com/articles/nphys2371} {\bibfield  {journal} {\bibinfo
  {journal} {Nature Physics}\ }\textbf {\bibinfo {volume} {8}},\ \bibinfo
  {pages} {662} (\bibinfo {year} {2012})}\BibitemShut {NoStop}%
\bibitem [{\citenamefont {Martens}\ \emph {et~al.}(2013)\citenamefont
  {Martens}, \citenamefont {Thutupalli}, \citenamefont {Fourri{\`e}re},\ and\
  \citenamefont {Hallatschek}}]{Martens10563}%
  \BibitemOpen
  \bibfield  {author} {\bibinfo {author} {\bibfnamefont {E.~A.}\ \bibnamefont
  {Martens}}, \bibinfo {author} {\bibfnamefont {S.}~\bibnamefont {Thutupalli}},
  \bibinfo {author} {\bibfnamefont {A.}~\bibnamefont {Fourri{\`e}re}},\ and\
  \bibinfo {author} {\bibfnamefont {O.}~\bibnamefont {Hallatschek}},\
  }\bibfield  {title} {\bibinfo {title} {Chimera states in mechanical
  oscillator networks},\ }\href {https://www.pnas.org/content/110/26/10563}
  {\bibfield  {journal} {\bibinfo  {journal} {Proceedings of the National
  Academy of Sciences}\ }\textbf {\bibinfo {volume} {110}},\ \bibinfo {pages}
  {10563} (\bibinfo {year} {2013})}\BibitemShut {NoStop}%
\bibitem [{\citenamefont {B\"ohm}\ \emph {et~al.}(2015)\citenamefont {B\"ohm},
  \citenamefont {Zakharova}, \citenamefont {Sch\"oll},\ and\ \citenamefont
  {L\"udge}}]{PhysRevE.91.040901}%
  \BibitemOpen
  \bibfield  {author} {\bibinfo {author} {\bibfnamefont {F.}~\bibnamefont
  {B\"ohm}}, \bibinfo {author} {\bibfnamefont {A.}~\bibnamefont {Zakharova}},
  \bibinfo {author} {\bibfnamefont {E.}~\bibnamefont {Sch\"oll}},\ and\
  \bibinfo {author} {\bibfnamefont {K.}~\bibnamefont {L\"udge}},\ }\bibfield
  {title} {\bibinfo {title} {Amplitude-phase coupling drives chimera states in
  globally coupled laser networks},\ }\href
  {https://doi.org/10.1103/PhysRevE.91.040901} {\bibfield  {journal} {\bibinfo
  {journal} {Phys. Rev. E}\ }\textbf {\bibinfo {volume} {91}},\ \bibinfo
  {pages} {040901} (\bibinfo {year} {2015})}\BibitemShut {NoStop}%
\bibitem [{\citenamefont {Larger}\ \emph {et~al.}(2015)\citenamefont {Larger},
  \citenamefont {Penkovsky},\ and\ \citenamefont
  {Maistrenko}}]{larger2015laser}%
  \BibitemOpen
  \bibfield  {author} {\bibinfo {author} {\bibfnamefont {L.}~\bibnamefont
  {Larger}}, \bibinfo {author} {\bibfnamefont {B.}~\bibnamefont {Penkovsky}},\
  and\ \bibinfo {author} {\bibfnamefont {Y.}~\bibnamefont {Maistrenko}},\
  }\bibfield  {title} {\bibinfo {title} {Laser chimeras as a paradigm for
  multistable patterns in complex systems},\ }\href@noop {} {\bibfield
  {journal} {\bibinfo  {journal} {Nature communications}\ }\textbf {\bibinfo
  {volume} {6}},\ \bibinfo {pages} {7752} (\bibinfo {year} {2015})}\BibitemShut
  {NoStop}%
\bibitem [{\citenamefont {Sethia}\ \emph {et~al.}(2013)\citenamefont {Sethia},
  \citenamefont {Sen},\ and\ \citenamefont {Johnston}}]{sethiasen1}%
  \BibitemOpen
  \bibfield  {author} {\bibinfo {author} {\bibfnamefont {G.~C.}\ \bibnamefont
  {Sethia}}, \bibinfo {author} {\bibfnamefont {A.}~\bibnamefont {Sen}},\ and\
  \bibinfo {author} {\bibfnamefont {G.~L.}\ \bibnamefont {Johnston}},\
  }\bibfield  {title} {\bibinfo {title} {Amplitude-mediated chimera states},\
  }\href {https://link.aps.org/doi/10.1103/PhysRevE.88.042917} {\bibfield
  {journal} {\bibinfo  {journal} {Phys. Rev. E}\ }\textbf {\bibinfo {volume}
  {88}},\ \bibinfo {pages} {042917} (\bibinfo {year} {2013})}\BibitemShut
  {NoStop}%
\bibitem [{\citenamefont {Schmidt}\ \emph {et~al.}(2014)\citenamefont
  {Schmidt}, \citenamefont {Schönleber}, \citenamefont {Krischer},\ and\
  \citenamefont {García-Morales}}]{schmidtglobal}%
  \BibitemOpen
  \bibfield  {author} {\bibinfo {author} {\bibfnamefont {L.}~\bibnamefont
  {Schmidt}}, \bibinfo {author} {\bibfnamefont {K.}~\bibnamefont
  {Schönleber}}, \bibinfo {author} {\bibfnamefont {K.}~\bibnamefont
  {Krischer}},\ and\ \bibinfo {author} {\bibfnamefont {V.}~\bibnamefont
  {García-Morales}},\ }\bibfield  {title} {\bibinfo {title} {Coexistence of
  synchrony and incoherence in oscillatory media under nonlinear global
  coupling},\ }\href {https://doi.org/10.1063/1.4858996} {\bibfield  {journal}
  {\bibinfo  {journal} {Chaos: An Interdisciplinary Journal of Nonlinear
  Science}\ }\textbf {\bibinfo {volume} {24}},\ \bibinfo {pages} {013102}
  (\bibinfo {year} {2014})}\BibitemShut {NoStop}%
\bibitem [{\citenamefont {Sethia}\ and\ \citenamefont {Sen}(2014)}]{sethiasen}%
  \BibitemOpen
  \bibfield  {author} {\bibinfo {author} {\bibfnamefont {G.~C.}\ \bibnamefont
  {Sethia}}\ and\ \bibinfo {author} {\bibfnamefont {A.}~\bibnamefont {Sen}},\
  }\bibfield  {title} {\bibinfo {title} {Chimera states: The existence criteria
  revisited},\ }\href {https://link.aps.org/doi/10.1103/PhysRevLett.112.144101}
  {\bibfield  {journal} {\bibinfo  {journal} {Phys. Rev. Lett.}\ }\textbf
  {\bibinfo {volume} {112}},\ \bibinfo {pages} {144101} (\bibinfo {year}
  {2014})}\BibitemShut {NoStop}%
\bibitem [{\citenamefont {Zakharova}\ \emph {et~al.}(2014)\citenamefont
  {Zakharova}, \citenamefont {Kapeller},\ and\ \citenamefont
  {Sch\"oll}}]{amplitudechimera}%
  \BibitemOpen
  \bibfield  {author} {\bibinfo {author} {\bibfnamefont {A.}~\bibnamefont
  {Zakharova}}, \bibinfo {author} {\bibfnamefont {M.}~\bibnamefont
  {Kapeller}},\ and\ \bibinfo {author} {\bibfnamefont {E.}~\bibnamefont
  {Sch\"oll}},\ }\bibfield  {title} {\bibinfo {title} {Chimera death: Symmetry
  breaking in dynamical networks},\ }\href
  {https://link.aps.org/doi/10.1103/PhysRevLett.112.154101} {\bibfield
  {journal} {\bibinfo  {journal} {Phys. Rev. Lett.}\ }\textbf {\bibinfo
  {volume} {112}},\ \bibinfo {pages} {154101} (\bibinfo {year}
  {2014})}\BibitemShut {NoStop}%
\bibitem [{\citenamefont {Nkomo}\ \emph {et~al.}(2013)\citenamefont {Nkomo},
  \citenamefont {Tinsley},\ and\ \citenamefont
  {Showalter}}]{chemicaltheoretical}%
  \BibitemOpen
  \bibfield  {author} {\bibinfo {author} {\bibfnamefont {S.}~\bibnamefont
  {Nkomo}}, \bibinfo {author} {\bibfnamefont {M.~R.}\ \bibnamefont {Tinsley}},\
  and\ \bibinfo {author} {\bibfnamefont {K.}~\bibnamefont {Showalter}},\
  }\bibfield  {title} {\bibinfo {title} {Chimera states in populations of
  nonlocally coupled chemical oscillators},\ }\href
  {https://link.aps.org/doi/10.1103/PhysRevLett.110.244102} {\bibfield
  {journal} {\bibinfo  {journal} {Phys. Rev. Lett.}\ }\textbf {\bibinfo
  {volume} {110}},\ \bibinfo {pages} {244102} (\bibinfo {year}
  {2013})}\BibitemShut {NoStop}%
\bibitem [{\citenamefont {Wickramasinghe}\ and\ \citenamefont
  {Kiss}(2013)}]{wickramasinghe2013spatially}%
  \BibitemOpen
  \bibfield  {author} {\bibinfo {author} {\bibfnamefont {M.}~\bibnamefont
  {Wickramasinghe}}\ and\ \bibinfo {author} {\bibfnamefont {I.~Z.}\
  \bibnamefont {Kiss}},\ }\bibfield  {title} {\bibinfo {title} {Spatially
  organized dynamical states in chemical oscillator networks: Synchronization,
  dynamical differentiation, and chimera patterns},\ }\href
  {https://doi.org/10.1371/journal.pone.0080586} {\bibfield  {journal}
  {\bibinfo  {journal} {PloS one}\ }\textbf {\bibinfo {volume} {8}},\ \bibinfo
  {pages} {e80586} (\bibinfo {year} {2013})}\BibitemShut {NoStop}%
\bibitem [{\citenamefont {Nkomo}\ \emph {et~al.}(2016)\citenamefont {Nkomo},
  \citenamefont {Tinsley},\ and\ \citenamefont {Showalter}}]{nkomo2016}%
  \BibitemOpen
  \bibfield  {author} {\bibinfo {author} {\bibfnamefont {S.}~\bibnamefont
  {Nkomo}}, \bibinfo {author} {\bibfnamefont {M.~R.}\ \bibnamefont {Tinsley}},\
  and\ \bibinfo {author} {\bibfnamefont {K.}~\bibnamefont {Showalter}},\
  }\bibfield  {title} {\bibinfo {title} {Chimera and chimera-like states in
  populations of nonlocally coupled homogeneous and heterogeneous chemical
  oscillators},\ }\href {https://doi.org/10.1063/1.4962631} {\bibfield
  {journal} {\bibinfo  {journal} {Chaos: An Interdisciplinary Journal of
  Nonlinear Science}\ }\textbf {\bibinfo {volume} {26}},\ \bibinfo {pages}
  {094826} (\bibinfo {year} {2016})}\BibitemShut {NoStop}%
\bibitem [{\citenamefont {Totz}\ \emph {et~al.}(2018)\citenamefont {Totz},
  \citenamefont {Rode}, \citenamefont {Tinsley}, \citenamefont {Showalter},\
  and\ \citenamefont {Engel}}]{totz2018spiral}%
  \BibitemOpen
  \bibfield  {author} {\bibinfo {author} {\bibfnamefont {J.~F.}\ \bibnamefont
  {Totz}}, \bibinfo {author} {\bibfnamefont {J.}~\bibnamefont {Rode}}, \bibinfo
  {author} {\bibfnamefont {M.~R.}\ \bibnamefont {Tinsley}}, \bibinfo {author}
  {\bibfnamefont {K.}~\bibnamefont {Showalter}},\ and\ \bibinfo {author}
  {\bibfnamefont {H.}~\bibnamefont {Engel}},\ }\bibfield  {title} {\bibinfo
  {title} {Spiral wave chimera states in large populations of coupled chemical
  oscillators},\ }\href {https://www.nature.com/articles/s41567-017-0005-8}
  {\bibfield  {journal} {\bibinfo  {journal} {Nature Physics}\ }\textbf
  {\bibinfo {volume} {14}},\ \bibinfo {pages} {282} (\bibinfo {year}
  {2018})}\BibitemShut {NoStop}%
\bibitem [{\citenamefont {Kumar}\ and\ \citenamefont
  {Gangopadhyay}(2022)}]{pkgg3}%
  \BibitemOpen
  \bibfield  {author} {\bibinfo {author} {\bibfnamefont {P.}~\bibnamefont
  {Kumar}}\ and\ \bibinfo {author} {\bibfnamefont {G.}~\bibnamefont
  {Gangopadhyay}},\ }\bibfield  {title} {\bibinfo {title} {Nonequilibrium
  thermodynamic characterization of chimeras in a continuum chemical oscillator
  system},\ }\href {https://link.aps.org/doi/10.1103/PhysRevE.105.034208}
  {\bibfield  {journal} {\bibinfo  {journal} {Phys. Rev. E}\ }\textbf {\bibinfo
  {volume} {105}},\ \bibinfo {pages} {034208} (\bibinfo {year}
  {2022})}\BibitemShut {NoStop}%
\bibitem [{\citenamefont {Omel'chenko}\ \emph {et~al.}(2008)\citenamefont
  {Omel'chenko}, \citenamefont {Maistrenko},\ and\ \citenamefont
  {Tass}}]{naturallink}%
  \BibitemOpen
  \bibfield  {author} {\bibinfo {author} {\bibfnamefont {O.~E.}\ \bibnamefont
  {Omel'chenko}}, \bibinfo {author} {\bibfnamefont {Y.~L.}\ \bibnamefont
  {Maistrenko}},\ and\ \bibinfo {author} {\bibfnamefont {P.~A.}\ \bibnamefont
  {Tass}},\ }\bibfield  {title} {\bibinfo {title} {Chimera states: The natural
  link between coherence and incoherence},\ }\href
  {https://link.aps.org/doi/10.1103/PhysRevLett.100.044105} {\bibfield
  {journal} {\bibinfo  {journal} {Phys. Rev. Lett.}\ }\textbf {\bibinfo
  {volume} {100}},\ \bibinfo {pages} {044105} (\bibinfo {year}
  {2008})}\BibitemShut {NoStop}%
\bibitem [{\citenamefont {Omelchenko}\ \emph {et~al.}(2011)\citenamefont
  {Omelchenko}, \citenamefont {Maistrenko}, \citenamefont {H\"ovel},\ and\
  \citenamefont {Sch\"oll}}]{coherenceincoherencetransition}%
  \BibitemOpen
  \bibfield  {author} {\bibinfo {author} {\bibfnamefont {I.}~\bibnamefont
  {Omelchenko}}, \bibinfo {author} {\bibfnamefont {Y.}~\bibnamefont
  {Maistrenko}}, \bibinfo {author} {\bibfnamefont {P.}~\bibnamefont
  {H\"ovel}},\ and\ \bibinfo {author} {\bibfnamefont {E.}~\bibnamefont
  {Sch\"oll}},\ }\bibfield  {title} {\bibinfo {title} {Loss of coherence in
  dynamical networks: Spatial chaos and chimera states},\ }\href
  {https://link.aps.org/doi/10.1103/PhysRevLett.106.234102} {\bibfield
  {journal} {\bibinfo  {journal} {Phys. Rev. Lett.}\ }\textbf {\bibinfo
  {volume} {106}},\ \bibinfo {pages} {234102} (\bibinfo {year}
  {2011})}\BibitemShut {NoStop}%
\bibitem [{\citenamefont {Schmidt}\ and\ \citenamefont
  {Krischer}(2015)}]{prerequisite}%
  \BibitemOpen
  \bibfield  {author} {\bibinfo {author} {\bibfnamefont {L.}~\bibnamefont
  {Schmidt}}\ and\ \bibinfo {author} {\bibfnamefont {K.}~\bibnamefont
  {Krischer}},\ }\bibfield  {title} {\bibinfo {title} {Clustering as a
  prerequisite for chimera states in globally coupled systems},\ }\href
  {https://link.aps.org/doi/10.1103/PhysRevLett.114.034101} {\bibfield
  {journal} {\bibinfo  {journal} {Phys. Rev. Lett.}\ }\textbf {\bibinfo
  {volume} {114}},\ \bibinfo {pages} {034101} (\bibinfo {year}
  {2015})}\BibitemShut {NoStop}%
\bibitem [{\citenamefont {Zhang}\ and\ \citenamefont
  {Motter}(2021)}]{strongchimera}%
  \BibitemOpen
  \bibfield  {author} {\bibinfo {author} {\bibfnamefont {Y.}~\bibnamefont
  {Zhang}}\ and\ \bibinfo {author} {\bibfnamefont {A.~E.}\ \bibnamefont
  {Motter}},\ }\bibfield  {title} {\bibinfo {title} {Mechanism for strong
  chimeras},\ }\href {https://link.aps.org/doi/10.1103/PhysRevLett.126.094101}
  {\bibfield  {journal} {\bibinfo  {journal} {Phys. Rev. Lett.}\ }\textbf
  {\bibinfo {volume} {126}},\ \bibinfo {pages} {094101} (\bibinfo {year}
  {2021})}\BibitemShut {NoStop}%
\bibitem [{\citenamefont {Schmidt}\ and\ \citenamefont
  {Krischer}(2014)}]{twocluster}%
  \BibitemOpen
  \bibfield  {author} {\bibinfo {author} {\bibfnamefont {L.}~\bibnamefont
  {Schmidt}}\ and\ \bibinfo {author} {\bibfnamefont {K.}~\bibnamefont
  {Krischer}},\ }\bibfield  {title} {\bibinfo {title} {Two-cluster solutions in
  an ensemble of generic limit-cycle oscillators with periodic self-forcing via
  the mean-field},\ }\href {https://doi.org/10.1103/PhysRevE.90.042911}
  {\bibfield  {journal} {\bibinfo  {journal} {Phys. Rev. E}\ }\textbf {\bibinfo
  {volume} {90}},\ \bibinfo {pages} {042911} (\bibinfo {year}
  {2014})}\BibitemShut {NoStop}%
\bibitem [{\citenamefont {Rao}\ and\ \citenamefont
  {Esposito}(2016)}]{Rao2016NonequilibriumThermodynamics}%
  \BibitemOpen
  \bibfield  {author} {\bibinfo {author} {\bibfnamefont {R.}~\bibnamefont
  {Rao}}\ and\ \bibinfo {author} {\bibfnamefont {M.}~\bibnamefont {Esposito}},\
  }\bibfield  {title} {\bibinfo {title} {Nonequilibrium thermodynamics of
  chemical reaction networks: Wisdom from stochastic thermodynamics},\ }\href
  {https://doi.org/10.1103/PhysRevX.6.041064} {\bibfield  {journal} {\bibinfo
  {journal} {Phys. Rev. X}\ }\textbf {\bibinfo {volume} {6}},\ \bibinfo {pages}
  {041064} (\bibinfo {year} {2016})}\BibitemShut {NoStop}%
\bibitem [{\citenamefont {Falasco}\ \emph {et~al.}(2018)\citenamefont
  {Falasco}, \citenamefont {Rao},\ and\ \citenamefont
  {Esposito}}]{Falasco2018InformationPatterns}%
  \BibitemOpen
  \bibfield  {author} {\bibinfo {author} {\bibfnamefont {G.}~\bibnamefont
  {Falasco}}, \bibinfo {author} {\bibfnamefont {R.}~\bibnamefont {Rao}},\ and\
  \bibinfo {author} {\bibfnamefont {M.}~\bibnamefont {Esposito}},\ }\bibfield
  {title} {\bibinfo {title} {{Information Thermodynamics of Turing Patterns}},\
  }\href {https://doi.org/10.1103/PhysRevLett.121.108301} {\bibfield  {journal}
  {\bibinfo  {journal} {Physical Review Letters}\ }\textbf {\bibinfo {volume}
  {121}},\ \bibinfo {pages} {108301} (\bibinfo {year} {2018})}\BibitemShut
  {NoStop}%
\bibitem [{\citenamefont {Cao}\ \emph {et~al.}(2015)\citenamefont {Cao},
  \citenamefont {Wang}, \citenamefont {Ouyang},\ and\ \citenamefont
  {Tu}}]{TUfree}%
  \BibitemOpen
  \bibfield  {author} {\bibinfo {author} {\bibfnamefont {Y.}~\bibnamefont
  {Cao}}, \bibinfo {author} {\bibfnamefont {H.}~\bibnamefont {Wang}}, \bibinfo
  {author} {\bibfnamefont {Q.}~\bibnamefont {Ouyang}},\ and\ \bibinfo {author}
  {\bibfnamefont {Y.}~\bibnamefont {Tu}},\ }\bibfield  {title} {\bibinfo
  {title} {The free-energy cost of accurate biochemical oscillations},\
  }\href@noop {} {\bibfield  {journal} {\bibinfo  {journal} {Nature physics}\
  }\textbf {\bibinfo {volume} {11}},\ \bibinfo {pages} {772} (\bibinfo {year}
  {2015})}\BibitemShut {NoStop}%
\bibitem [{\citenamefont {Oberreiter}\ \emph {et~al.}(2022)\citenamefont
  {Oberreiter}, \citenamefont {Seifert},\ and\ \citenamefont
  {Barato}}]{barato}%
  \BibitemOpen
  \bibfield  {author} {\bibinfo {author} {\bibfnamefont {L.}~\bibnamefont
  {Oberreiter}}, \bibinfo {author} {\bibfnamefont {U.}~\bibnamefont
  {Seifert}},\ and\ \bibinfo {author} {\bibfnamefont {A.~C.}\ \bibnamefont
  {Barato}},\ }\bibfield  {title} {\bibinfo {title} {Universal minimal cost of
  coherent biochemical oscillations},\ }\href
  {https://link.aps.org/doi/10.1103/PhysRevE.106.014106} {\bibfield  {journal}
  {\bibinfo  {journal} {Phys. Rev. E}\ }\textbf {\bibinfo {volume} {106}},\
  \bibinfo {pages} {014106} (\bibinfo {year} {2022})}\BibitemShut {NoStop}%
\bibitem [{\citenamefont {Tufillaro}\ \emph {et~al.}(1989)\citenamefont
  {Tufillaro}, \citenamefont {Ramshankar},\ and\ \citenamefont
  {Gollub}}]{mainFaraday}%
  \BibitemOpen
  \bibfield  {author} {\bibinfo {author} {\bibfnamefont {N.~B.}\ \bibnamefont
  {Tufillaro}}, \bibinfo {author} {\bibfnamefont {R.}~\bibnamefont
  {Ramshankar}},\ and\ \bibinfo {author} {\bibfnamefont {J.~P.}\ \bibnamefont
  {Gollub}},\ }\bibfield  {title} {\bibinfo {title} {Order-disorder transition
  in capillary ripples},\ }\href
  {https://link.aps.org/doi/10.1103/PhysRevLett.62.422} {\bibfield  {journal}
  {\bibinfo  {journal} {Phys. Rev. Lett.}\ }\textbf {\bibinfo {volume} {62}},\
  \bibinfo {pages} {422} (\bibinfo {year} {1989})}\BibitemShut {NoStop}%
\bibitem [{\citenamefont {Braun}\ \emph {et~al.}(1991)\citenamefont {Braun},
  \citenamefont {Rasenat},\ and\ \citenamefont {Steinberg}}]{EBraun_1991}%
  \BibitemOpen
  \bibfield  {author} {\bibinfo {author} {\bibfnamefont {E.}~\bibnamefont
  {Braun}}, \bibinfo {author} {\bibfnamefont {S.}~\bibnamefont {Rasenat}},\
  and\ \bibinfo {author} {\bibfnamefont {V.}~\bibnamefont {Steinberg}},\
  }\bibfield  {title} {\bibinfo {title} {Mechanism of transition to a weak
  turbulence in extended anisotropic systems},\ }\href
  {https://dx.doi.org/10.1209/0295-5075/15/6/006} {\bibfield  {journal}
  {\bibinfo  {journal} {Europhysics Letters}\ }\textbf {\bibinfo {volume}
  {15}},\ \bibinfo {pages} {597} (\bibinfo {year} {1991})}\BibitemShut
  {NoStop}%
\bibitem [{\citenamefont {Goldman}\ \emph {et~al.}(2003)\citenamefont
  {Goldman}, \citenamefont {Shattuck}, \citenamefont {Moon}, \citenamefont
  {Swift},\ and\ \citenamefont {Swinney}}]{Granulargas}%
  \BibitemOpen
  \bibfield  {author} {\bibinfo {author} {\bibfnamefont {D.~I.}\ \bibnamefont
  {Goldman}}, \bibinfo {author} {\bibfnamefont {M.~D.}\ \bibnamefont
  {Shattuck}}, \bibinfo {author} {\bibfnamefont {S.~J.}\ \bibnamefont {Moon}},
  \bibinfo {author} {\bibfnamefont {J.~B.}\ \bibnamefont {Swift}},\ and\
  \bibinfo {author} {\bibfnamefont {H.~L.}\ \bibnamefont {Swinney}},\
  }\bibfield  {title} {\bibinfo {title} {Lattice dynamics and melting of a
  nonequilibrium pattern},\ }\href
  {https://link.aps.org/doi/10.1103/PhysRevLett.90.104302} {\bibfield
  {journal} {\bibinfo  {journal} {Phys. Rev. Lett.}\ }\textbf {\bibinfo
  {volume} {90}},\ \bibinfo {pages} {104302} (\bibinfo {year}
  {2003})}\BibitemShut {NoStop}%
\bibitem [{\citenamefont {Miethe}\ \emph {et~al.}(2009)\citenamefont {Miethe},
  \citenamefont {Garc\'{\i}a-Morales},\ and\ \citenamefont
  {Krischer}}]{mcgle1st}%
  \BibitemOpen
  \bibfield  {author} {\bibinfo {author} {\bibfnamefont {I.}~\bibnamefont
  {Miethe}}, \bibinfo {author} {\bibfnamefont {V.}~\bibnamefont
  {Garc\'{\i}a-Morales}},\ and\ \bibinfo {author} {\bibfnamefont
  {K.}~\bibnamefont {Krischer}},\ }\bibfield  {title} {\bibinfo {title}
  {Irregular subharmonic cluster patterns in an autonomous photoelectrochemical
  oscillator},\ }\href {https://doi.org/10.1103/PhysRevLett.102.194101}
  {\bibfield  {journal} {\bibinfo  {journal} {Phys. Rev. Lett.}\ }\textbf
  {\bibinfo {volume} {102}},\ \bibinfo {pages} {194101} (\bibinfo {year}
  {2009})}\BibitemShut {NoStop}%
\bibitem [{\citenamefont {Garc\'{\i}a-Morales}\ \emph
  {et~al.}(2010)\citenamefont {Garc\'{\i}a-Morales}, \citenamefont {Orlov},\
  and\ \citenamefont {Krischer}}]{mcgle2}%
  \BibitemOpen
  \bibfield  {author} {\bibinfo {author} {\bibfnamefont {V.}~\bibnamefont
  {Garc\'{\i}a-Morales}}, \bibinfo {author} {\bibfnamefont {A.}~\bibnamefont
  {Orlov}},\ and\ \bibinfo {author} {\bibfnamefont {K.}~\bibnamefont
  {Krischer}},\ }\bibfield  {title} {\bibinfo {title} {Subharmonic phase
  clusters in the complex ginzburg-landau equation with nonlinear global
  coupling},\ }\href {https://doi.org/10.1103/PhysRevE.82.065202} {\bibfield
  {journal} {\bibinfo  {journal} {Phys. Rev. E}\ }\textbf {\bibinfo {volume}
  {82}},\ \bibinfo {pages} {065202} (\bibinfo {year} {2010})}\BibitemShut
  {NoStop}%
\bibitem [{\citenamefont {Prigogine}\ and\ \citenamefont
  {Lefever}(1968)}]{Prigogine1968SymmetryII}%
  \BibitemOpen
  \bibfield  {author} {\bibinfo {author} {\bibfnamefont {I.}~\bibnamefont
  {Prigogine}}\ and\ \bibinfo {author} {\bibfnamefont {R.}~\bibnamefont
  {Lefever}},\ }\bibfield  {title} {\bibinfo {title} {{Symmetry Breaking
  Instabilities in Dissipative Systems. II}},\ }\href
  {https://doi.org/10.1063/1.1668896} {\bibfield  {journal} {\bibinfo
  {journal} {The Journal of Chemical Physics}\ }\textbf {\bibinfo {volume}
  {48}},\ \bibinfo {pages} {1695} (\bibinfo {year} {1968})}\BibitemShut
  {NoStop}%
\bibitem [{\citenamefont {Nicolis}\ and\ \citenamefont
  {Prigogine}(1977)}]{Nicolis1977Self-organizationFluctuations}%
  \BibitemOpen
  \bibfield  {author} {\bibinfo {author} {\bibfnamefont {G.}~\bibnamefont
  {Nicolis}}\ and\ \bibinfo {author} {\bibfnamefont {I.~I.}\ \bibnamefont
  {Prigogine}},\ }\href
  {https://books.google.co.in/books/about/Self_Organization_in_Nonequilibrium_Syst.html?id=mZkQAQAAIAAJ&redir_esc=y}
  {\emph {\bibinfo {title} {{Self-organization in nonequilibrium systems : from
  dissipative structures to order through fluctuations}}}}\ (\bibinfo
  {publisher} {Wiley},\ \bibinfo {year} {1977})\ p.\ \bibinfo {pages}
  {491}\BibitemShut {NoStop}%
\bibitem [{\citenamefont {Nicolis}(1995)}]{Nicolis1995IntroductionScience}%
  \BibitemOpen
  \bibfield  {author} {\bibinfo {author} {\bibfnamefont {G.}~\bibnamefont
  {Nicolis}},\ }\href@noop {} {\emph {\bibinfo {title} {{Introduction to
  nonlinear science}}}}\ (\bibinfo  {publisher} {Cambridge University Press},\
  \bibinfo {year} {1995})\ p.\ \bibinfo {pages} {254}\BibitemShut {NoStop}%
\bibitem [{\citenamefont {Cross}\ and\ \citenamefont
  {Greenside}(2009)}]{Cross2009PatternSystems}%
  \BibitemOpen
  \bibfield  {author} {\bibinfo {author} {\bibfnamefont {M.}~\bibnamefont
  {Cross}}\ and\ \bibinfo {author} {\bibfnamefont {H.}~\bibnamefont
  {Greenside}},\ }\href {http://ebooks.cambridge.org/ref/id/CBO9780511627200}
  {\emph {\bibinfo {title} {{Pattern Formation and Dynamics in Nonequilibrium
  Systems}}}}\ (\bibinfo  {publisher} {Cambridge University Press},\ \bibinfo
  {address} {Cambridge},\ \bibinfo {year} {2009})\BibitemShut {NoStop}%
\bibitem [{\citenamefont {Krylov}\ and\ \citenamefont
  {Bogoliubov}(1949)}]{krylov1949introduction}%
  \BibitemOpen
  \bibfield  {author} {\bibinfo {author} {\bibfnamefont {N.~M.}\ \bibnamefont
  {Krylov}}\ and\ \bibinfo {author} {\bibfnamefont {N.~N.}\ \bibnamefont
  {Bogoliubov}},\ }\href@noop {} {\emph {\bibinfo {title} {Introduction to
  non-linear mechanics}}}\ (\bibinfo  {publisher} {Princeton University
  Press},\ \bibinfo {year} {1949})\BibitemShut {NoStop}%
\bibitem [{\citenamefont {Kumar}\ and\ \citenamefont
  {Gangopadhyay}(2020)}]{pkgg}%
  \BibitemOpen
  \bibfield  {author} {\bibinfo {author} {\bibfnamefont {P.}~\bibnamefont
  {Kumar}}\ and\ \bibinfo {author} {\bibfnamefont {G.}~\bibnamefont
  {Gangopadhyay}},\ }\bibfield  {title} {\bibinfo {title} {Energetic and
  entropic cost due to overlapping of turing-hopf instabilities in the presence
  of cross diffusion},\ }\href {https://doi.org/10.1103/PhysRevE.101.042204}
  {\bibfield  {journal} {\bibinfo  {journal} {Phys. Rev. E}\ }\textbf {\bibinfo
  {volume} {101}},\ \bibinfo {pages} {042204} (\bibinfo {year}
  {2020})}\BibitemShut {NoStop}%
\bibitem [{\citenamefont {Cox}\ and\ \citenamefont {Matthews}(2002)}]{COX2002}%
  \BibitemOpen
  \bibfield  {author} {\bibinfo {author} {\bibfnamefont {S.}~\bibnamefont
  {Cox}}\ and\ \bibinfo {author} {\bibfnamefont {P.}~\bibnamefont {Matthews}},\
  }\bibfield  {title} {\bibinfo {title} {Exponential time differencing for
  stiff systems},\ }\href
  {https://doi.org/https://doi.org/10.1006/jcph.2002.6995} {\bibfield
  {journal} {\bibinfo  {journal} {Journal of Computational Physics}\ }\textbf
  {\bibinfo {volume} {176}},\ \bibinfo {pages} {430} (\bibinfo {year}
  {2002})}\BibitemShut {NoStop}%
\bibitem [{\citenamefont {Prigogine}(1954)}]{Prigogine1954ChemicalDefay.}%
  \BibitemOpen
  \bibfield  {author} {\bibinfo {author} {\bibfnamefont {I.}~\bibnamefont
  {Prigogine}},\ }\href
  {https://www.worldcat.org/title/chemical-thermodynamics-by-i-prigogine-and-r-defay/oclc/61574676}
  {\emph {\bibinfo {title} {{Chemical thermodynamics [by] I. Prigogine and R.
  Defay.}}}}\ (\bibinfo  {publisher} {Longmans},\ \bibinfo {address}
  {[London]},\ \bibinfo {year} {1954})\BibitemShut {NoStop}%
\bibitem [{\citenamefont {Kondepudi}\ and\ \citenamefont
  {Prigogine}(2014)}]{Kondepudi2014ModernThermodynamics}%
  \BibitemOpen
  \bibfield  {author} {\bibinfo {author} {\bibfnamefont {D.}~\bibnamefont
  {Kondepudi}}\ and\ \bibinfo {author} {\bibfnamefont {I.}~\bibnamefont
  {Prigogine}},\ }\href {https://doi.org/10.1002/9781118698723} {\emph
  {\bibinfo {title} {{Modern Thermodynamics}}}}\ (\bibinfo  {publisher} {John
  Wiley {\&} Sons, Ltd},\ \bibinfo {address} {Chichester, UK},\ \bibinfo {year}
  {2014})\BibitemShut {NoStop}%
\bibitem [{\citenamefont {Alberty}(2003)}]{Alberty2003ThermodynamicsReactions}%
  \BibitemOpen
  \bibfield  {author} {\bibinfo {author} {\bibfnamefont {R.~A.}\ \bibnamefont
  {Alberty}},\ }\href {https://doi.org/10.1002/0471332607} {\emph {\bibinfo
  {title} {{Thermodynamics of Biochemical Reactions}}}}\ (\bibinfo  {publisher}
  {John Wiley {\&} Sons, Inc.},\ \bibinfo {address} {Hoboken, NJ, USA},\
  \bibinfo {year} {2003})\BibitemShut {NoStop}%
\bibitem [{\citenamefont {Rao}\ and\ \citenamefont
  {Esposito}(2018)}]{Raoconservationlaw}%
  \BibitemOpen
  \bibfield  {author} {\bibinfo {author} {\bibfnamefont {R.}~\bibnamefont
  {Rao}}\ and\ \bibinfo {author} {\bibfnamefont {M.}~\bibnamefont {Esposito}},\
  }\bibfield  {title} {\bibinfo {title} {Conservation laws and work fluctuation
  relations in chemical reaction networks},\ }\href
  {https://doi.org/10.1063/1.5042253} {\bibfield  {journal} {\bibinfo
  {journal} {The Journal of Chemical Physics}\ }\textbf {\bibinfo {volume}
  {149}},\ \bibinfo {pages} {245101} (\bibinfo {year} {2018})}\BibitemShut
  {NoStop}%
\bibitem [{\citenamefont {Cover}\ and\ \citenamefont
  {Thomas}(1999)}]{Cover1999ElementsTheory}%
  \BibitemOpen
  \bibfield  {author} {\bibinfo {author} {\bibfnamefont {T.~M.}\ \bibnamefont
  {Cover}}\ and\ \bibinfo {author} {\bibfnamefont {J.~A.}\ \bibnamefont
  {Thomas}},\ }\href@noop {} {\emph {\bibinfo {title} {{Elements of information
  theory}}}}\ (\bibinfo  {publisher} {Wiley-India},\ \bibinfo {year} {1999})\
  p.\ \bibinfo {pages} {542}\BibitemShut {NoStop}%
\bibitem [{\citenamefont {Avanzini}\ \emph {et~al.}(2019)\citenamefont
  {Avanzini}, \citenamefont {Falasco},\ and\ \citenamefont
  {Esposito}}]{thermodynamicschemifcalwaves}%
  \BibitemOpen
  \bibfield  {author} {\bibinfo {author} {\bibfnamefont {F.}~\bibnamefont
  {Avanzini}}, \bibinfo {author} {\bibfnamefont {G.}~\bibnamefont {Falasco}},\
  and\ \bibinfo {author} {\bibfnamefont {M.}~\bibnamefont {Esposito}},\
  }\bibfield  {title} {\bibinfo {title} {Thermodynamics of chemical waves},\
  }\href {https://doi.org/10.1063/1.5126528} {\bibfield  {journal} {\bibinfo
  {journal} {The Journal of Chemical Physics}\ }\textbf {\bibinfo {volume}
  {151}},\ \bibinfo {pages} {234103} (\bibinfo {year} {2019})},\ \Eprint
  {https://arxiv.org/abs/https://doi.org/10.1063/1.5126528}
  {https://doi.org/10.1063/1.5126528} \BibitemShut {NoStop}%
\bibitem [{\citenamefont {Gopal}\ \emph {et~al.}(2014)\citenamefont {Gopal},
  \citenamefont {Chandrasekar}, \citenamefont {Venkatesan},\ and\ \citenamefont
  {Lakshmanan}}]{laks}%
  \BibitemOpen
  \bibfield  {author} {\bibinfo {author} {\bibfnamefont {R.}~\bibnamefont
  {Gopal}}, \bibinfo {author} {\bibfnamefont {V.~K.}\ \bibnamefont
  {Chandrasekar}}, \bibinfo {author} {\bibfnamefont {A.}~\bibnamefont
  {Venkatesan}},\ and\ \bibinfo {author} {\bibfnamefont {M.}~\bibnamefont
  {Lakshmanan}},\ }\bibfield  {title} {\bibinfo {title} {Observation and
  characterization of chimera states in coupled dynamical systems with nonlocal
  coupling},\ }\href {https://link.aps.org/doi/10.1103/PhysRevE.89.052914}
  {\bibfield  {journal} {\bibinfo  {journal} {Phys. Rev. E}\ }\textbf {\bibinfo
  {volume} {89}},\ \bibinfo {pages} {052914} (\bibinfo {year}
  {2014})}\BibitemShut {NoStop}%
\bibitem [{\citenamefont {Zakharova}(2020)}]{zakharova2020chimera}%
  \BibitemOpen
  \bibfield  {author} {\bibinfo {author} {\bibfnamefont {A.}~\bibnamefont
  {Zakharova}},\ }\href@noop {} {\emph {\bibinfo {title} {Chimera Patterns in
  Networks}}}\ (\bibinfo  {publisher} {Springer},\ \bibinfo {year}
  {2020})\BibitemShut {NoStop}%
\bibitem [{\citenamefont {Omelchenko}\ \emph {et~al.}(2013)\citenamefont
  {Omelchenko}, \citenamefont {Omel'chenko}, \citenamefont {H\"ovel},\ and\
  \citenamefont {Sch\"oll}}]{multichimera}%
  \BibitemOpen
  \bibfield  {author} {\bibinfo {author} {\bibfnamefont {I.}~\bibnamefont
  {Omelchenko}}, \bibinfo {author} {\bibfnamefont {O.~E.}\ \bibnamefont
  {Omel'chenko}}, \bibinfo {author} {\bibfnamefont {P.}~\bibnamefont
  {H\"ovel}},\ and\ \bibinfo {author} {\bibfnamefont {E.}~\bibnamefont
  {Sch\"oll}},\ }\bibfield  {title} {\bibinfo {title} {When nonlocal coupling
  between oscillators becomes stronger: Patched synchrony or multichimera
  states},\ }\href {https://link.aps.org/doi/10.1103/PhysRevLett.110.224101}
  {\bibfield  {journal} {\bibinfo  {journal} {Phys. Rev. Lett.}\ }\textbf
  {\bibinfo {volume} {110}},\ \bibinfo {pages} {224101} (\bibinfo {year}
  {2013})}\BibitemShut {NoStop}%
\bibitem [{\citenamefont {Lemoult}\ \emph {et~al.}(2016)\citenamefont
  {Lemoult}, \citenamefont {Shi}, \citenamefont {Avila}, \citenamefont
  {Jalikop}, \citenamefont {Avila},\ and\ \citenamefont {Hof}}]{Couetteflow}%
  \BibitemOpen
  \bibfield  {author} {\bibinfo {author} {\bibfnamefont {G.}~\bibnamefont
  {Lemoult}}, \bibinfo {author} {\bibfnamefont {L.}~\bibnamefont {Shi}},
  \bibinfo {author} {\bibfnamefont {K.}~\bibnamefont {Avila}}, \bibinfo
  {author} {\bibfnamefont {S.~V.}\ \bibnamefont {Jalikop}}, \bibinfo {author}
  {\bibfnamefont {M.}~\bibnamefont {Avila}},\ and\ \bibinfo {author}
  {\bibfnamefont {B.}~\bibnamefont {Hof}},\ }\bibfield  {title} {\bibinfo
  {title} {Directed percolation phase transition to sustained turbulence in
  couette flow},\ }\href {https://www.nature.com/articles/nphys3675} {\bibfield
   {journal} {\bibinfo  {journal} {Nature Physics}\ }\textbf {\bibinfo {volume}
  {12}},\ \bibinfo {pages} {254} (\bibinfo {year} {2016})}\BibitemShut
  {NoStop}%
\bibitem [{\citenamefont {Chat{\'e}}\ and\ \citenamefont
  {Manneville}(1987)}]{chate1987transition}%
  \BibitemOpen
  \bibfield  {author} {\bibinfo {author} {\bibfnamefont {H.}~\bibnamefont
  {Chat{\'e}}}\ and\ \bibinfo {author} {\bibfnamefont {P.}~\bibnamefont
  {Manneville}},\ }\bibfield  {title} {\bibinfo {title} {Transition to
  turbulence via spatio-temporal intermittency},\ }\href
  {https://link.aps.org/doi/10.1103/PhysRevLett.58.112} {\bibfield  {journal}
  {\bibinfo  {journal} {Phys. Rev. Lett.}\ }\textbf {\bibinfo {volume} {58}},\
  \bibinfo {pages} {112} (\bibinfo {year} {1987})}\BibitemShut {NoStop}%
\bibitem [{\citenamefont {Kaneko}(1985)}]{kaneko1985spatiotemporal}%
  \BibitemOpen
  \bibfield  {author} {\bibinfo {author} {\bibfnamefont {K.}~\bibnamefont
  {Kaneko}},\ }\bibfield  {title} {\bibinfo {title} {Spatiotemporal
  intermittency in coupled map lattices},\ }\href
  {https://doi.org/10.1143/PTP.74.1033} {\bibfield  {journal} {\bibinfo
  {journal} {Progress of Theoretical Physics}\ }\textbf {\bibinfo {volume}
  {74}},\ \bibinfo {pages} {1033} (\bibinfo {year} {1985})}\BibitemShut
  {NoStop}%
\bibitem [{\citenamefont {Omel'chenko}\ \emph {et~al.}(2010)\citenamefont
  {Omel'chenko}, \citenamefont {Wolfrum},\ and\ \citenamefont
  {Maistrenko}}]{chaoticspatiotemporal}%
  \BibitemOpen
  \bibfield  {author} {\bibinfo {author} {\bibfnamefont {O.~E.}\ \bibnamefont
  {Omel'chenko}}, \bibinfo {author} {\bibfnamefont {M.}~\bibnamefont
  {Wolfrum}},\ and\ \bibinfo {author} {\bibfnamefont {Y.~L.}\ \bibnamefont
  {Maistrenko}},\ }\bibfield  {title} {\bibinfo {title} {Chimera states as
  chaotic spatiotemporal patterns},\ }\href
  {https://link.aps.org/doi/10.1103/PhysRevE.81.065201} {\bibfield  {journal}
  {\bibinfo  {journal} {Phys. Rev. E}\ }\textbf {\bibinfo {volume} {81}},\
  \bibinfo {pages} {065201} (\bibinfo {year} {2010})}\BibitemShut {NoStop}%
\bibitem [{\citenamefont {Nakagawa}\ and\ \citenamefont
  {Kuramoto}(1993)}]{NakagawaKuramoto}%
  \BibitemOpen
  \bibfield  {author} {\bibinfo {author} {\bibfnamefont {N.}~\bibnamefont
  {Nakagawa}}\ and\ \bibinfo {author} {\bibfnamefont {Y.}~\bibnamefont
  {Kuramoto}},\ }\bibfield  {title} {\bibinfo {title} {{Collective Chaos in a
  Population of Globally Coupled Oscillators}},\ }\href
  {https://doi.org/10.1143/ptp/89.2.313} {\bibfield  {journal} {\bibinfo
  {journal} {Progress of Theoretical Physics}\ }\textbf {\bibinfo {volume}
  {89}},\ \bibinfo {pages} {313} (\bibinfo {year} {1993})}\BibitemShut
  {NoStop}%
\bibitem [{\citenamefont {Kotwal}\ \emph {et~al.}(2017)\citenamefont {Kotwal},
  \citenamefont {Jiang},\ and\ \citenamefont {Abrams}}]{Kotwal}%
  \BibitemOpen
  \bibfield  {author} {\bibinfo {author} {\bibfnamefont {T.}~\bibnamefont
  {Kotwal}}, \bibinfo {author} {\bibfnamefont {X.}~\bibnamefont {Jiang}},\ and\
  \bibinfo {author} {\bibfnamefont {D.~M.}\ \bibnamefont {Abrams}},\ }\bibfield
   {title} {\bibinfo {title} {Connecting the kuramoto model and the chimera
  state},\ }\href {https://doi.org/10.1103/PhysRevLett.119.264101} {\bibfield
  {journal} {\bibinfo  {journal} {Phys. Rev. Lett.}\ }\textbf {\bibinfo
  {volume} {119}},\ \bibinfo {pages} {264101} (\bibinfo {year}
  {2017})}\BibitemShut {NoStop}%
\bibitem [{\citenamefont {Kumar}\ and\ \citenamefont
  {Gangopadhyay}(2021)}]{pkgg2}%
  \BibitemOpen
  \bibfield  {author} {\bibinfo {author} {\bibfnamefont {P.}~\bibnamefont
  {Kumar}}\ and\ \bibinfo {author} {\bibfnamefont {G.}~\bibnamefont
  {Gangopadhyay}},\ }\bibfield  {title} {\bibinfo {title} {Nonequilibrium
  thermodynamics of glycolytic traveling wave: Benjamin-feir instability},\
  }\href {https://doi.org/10.1103/PhysRevE.104.014221} {\bibfield  {journal}
  {\bibinfo  {journal} {Phys. Rev. E}\ }\textbf {\bibinfo {volume} {104}},\
  \bibinfo {pages} {014221} (\bibinfo {year} {2021})}\BibitemShut {NoStop}%
\bibitem [{\citenamefont {Bansal}\ \emph {et~al.}(2019)\citenamefont {Bansal},
  \citenamefont {Garcia}, \citenamefont {Tompson}, \citenamefont {Verstynen},
  \citenamefont {Vettel},\ and\ \citenamefont {Muldoon}}]{kanikabrain}%
  \BibitemOpen
  \bibfield  {author} {\bibinfo {author} {\bibfnamefont {K.}~\bibnamefont
  {Bansal}}, \bibinfo {author} {\bibfnamefont {J.~O.}\ \bibnamefont {Garcia}},
  \bibinfo {author} {\bibfnamefont {S.~H.}\ \bibnamefont {Tompson}}, \bibinfo
  {author} {\bibfnamefont {T.}~\bibnamefont {Verstynen}}, \bibinfo {author}
  {\bibfnamefont {J.~M.}\ \bibnamefont {Vettel}},\ and\ \bibinfo {author}
  {\bibfnamefont {S.~F.}\ \bibnamefont {Muldoon}},\ }\bibfield  {title}
  {\bibinfo {title} {Cognitive chimera states in human brain networks},\ }\href
  {https://www.science.org/doi/abs/10.1126/sciadv.aau8535} {\bibfield
  {journal} {\bibinfo  {journal} {Science Advances}\ }\textbf {\bibinfo
  {volume} {5}},\ \bibinfo {pages} {eaau8535} (\bibinfo {year}
  {2019})}\BibitemShut {NoStop}%
\end{thebibliography}%
\end{document}